\documentclass[twocolumn]{aastex631}

\begin{document}

\newcommand{\msun}{{M_\odot}}
\newcommand{\rsun}{{R_\odot}}
\newcommand{\lsun}{{L_\odot}}
\newcommand{\tess}{{TESS}}
\newcommand{\gaia}{{\em Gaia}}
\newcommand{\rapid}{{{\tt rapid}}}
\newcommand{\delayed}{{{\tt delayed}}}
\newcommand{\porb}{{{P_{\rm{orb}}}}}
\newcommand{\metal}{{{\rm Z}}}

\newcommand{\mlc}{{{M_{\rm{LC}}}}}
\newcommand{\llc}{{{L_{\rm{LC}}}}}
\newcommand{\myr}{{{\rm Myr}}}
\newcommand{\mbh}{{{M_{\rm{BH}}}}}
\newcommand{\mns}{{{M_{\rm{NS}}}}}
\newcommand{\mwd}{{{M_{\rm{WD}}}}}
\newcommand{\mco}{{{M_{\rm{CO}}}}}

\newcommand{\mtot}{{M_{\rm{tot}}}}
\newcommand{\ecc}{{Ecc}}
\newcommand{\yr}{{{\rm{yr}}}}
\newcommand{\gyr}{{{\rm{Gyr}}}}

\newcommand{\kpc}{{{\rm{kpc}}}}
\newcommand{\pc}{{{\rm{pc}}}}
\newcommand{\BH}{{\rm{BH}}}
\newcommand{\LC}{{\rm{LC}}}
\newcommand{\cosmic}{{{\texttt{COSMIC}}}}
\newcommand{\mwdust}{{\texttt{mwdust}}}
\newcommand{\sse}{{{\texttt{SSE}}}}
\newcommand{\bse}{{{\texttt{BSE}}}}
\newcommand{\Mmin}{{M_{\rm{min}}}}
\newcommand{\Mmax}{{M_{\rm{max}}}}
\newcommand{\sma}{{a}}
\newcommand{\fallback}{{\rm{FB}}}
\newcommand{\bhlc}{{\rm{BH\text{--}LC}}}
\newcommand{\nslc}{{\rm{NS\text{--}LC}}}
\newcommand{\wdlc}{{\rm{WD\text{--}LC}}}
\newcommand{\colc}{{\rm{CO\text{--}LC}}}
\newcommand{\mtwelve}{{\bf{m12i}}}
\newcommand{\beam}{{{\rm{RB}}}}
\newcommand{\kms}{{\rm{km\ s^{-1}}}}
\newcommand{\tdur}{{t_{\rm{dur}}}}
\newcommand{\snr}{{{\rm{SNR}}}}
\newcommand{\days}{{\rm{day}}}
\newcommand{\ergspers}{{{\rm{ergs\ s^{-1}}}}}
\newcommand{\macc}{{M_{\rm{acc}}}}
\newcommand{\racc}{{R_{\rm{acc}}}}
\newcommand{\mzams}{{M_{\rm{ZAMS}}}}
\newcommand{\rzams}{{R_{\rm{ZAMS}}}}
\newcommand{\mch}{{M_{\rm{Ch}}}}
\newcommand{\qmin}{{q_\mathrm{min}}}
\newcommand{\nss}{{\em NSS}}
\newcommand{\mprim}{{m_{\rm{pri}}}}
\newcommand{\msec}{{m_{\rm{sec}}}}
\newcommand{\rpri}{{R_{\rm{pri}}}}
\newcommand{\rsec}{{R_{\rm{sec}}}}
\newcommand{\rRoche}{{R_{\rm{RL}}}}
\newcommand{\kmps}{{\rm{km\ s^{-1}}}}
\newcommand{\match}{{\rm{Match}}}
\newcommand{\ncolcmw}{{ N_{\colc, \rm{MW} } }}
\newcommand{\ncolcsim}{{ N_{\colc, \rm{sim} } }}
\newcommand{\sourav}[1]{\textbf{\color{red} Sourav: #1}}
\newcommand{\kb}[1]{\textbf{\color{teal} KB: #1}}
\newcommand{\cc}[1]{\textbf{\color{blue} CC: #1}}

\title{\gaia's promise to detect compact-object binaries: where we stand with the third data release}

\author[0000-0001-9685-3777]{Chirag Chawla}
\altaffiliation{IISER-Rubin Observatory Fellow}
\affiliation{Department of Physics, Indian Institute of Science Education and Research, Dr. Homi Bhabha Road, Pune, 411008, India}
\affiliation{Department of Astronomy and Astrophysics, Tata Institute of Fundamental Research, Homi Bhabha Road, Navy Nagar, Colaba, Mumbai, 400005, India}
\email{chiragchawla134@gmail.com}

\author[0000-0002-3680-2684]{Sourav Chatterjee}
\affiliation{Department of Astronomy and Astrophysics, Tata Institute of Fundamental Research, Homi Bhabha Road, Navy Nagar, Colaba, Mumbai, 400005, India}
\email{souravchatterjee.tifr@gmail.com}

\author[0000-0001-5228-6598]{Katelyn Breivik}
\affiliation{Department of Physics, McWilliams Center for Cosmology and Astrophysics, Carnegie Mellon University, Pittsburgh, PA 15213, USA}

\begin{abstract}
\label{sec:abstract}
With its third data release (DR3), \gaia\ begins unveiling dormant candidate compact object (CO) binaries with luminous companions (LC) as predicted by several past theoretical studies. To date, 3 black hole (BH), 21 neutron star (NS), and $\approx3200$ white dwarf (WD) candidates have been identified with LCs in detached orbits using astrometry. We adopt an observationally motivated sampling scheme for the star formation history of the Milky Way, and initial zero-age main-sequence binary properties, incorporate all relevant binary interaction processes during evolution to obtain a realistic present-day intrinsic population of \colc\ binaries. We apply \gaia's selection criteria to identify the \colc\ binaries detectable using the observational cuts applicable for DR3 as well as its end-of-mission (EOM). We find that under the DR3 selection cuts, our detectable population includes no BH--LCs, approximately $10$--$40$ NS--LCs, and around $\sim4300$ WD--LCs. Our predicted NS--LC population is in good agreement with the current DR3 census, both in its predicted yield and in the orbital and stellar properties, and we recover a close analogue of the Gaia NS1 candidate together with its detailed formation pathway. For WD--LCs, we find that a moderate natal kick of $5$--$15\,\kms$ imparted at WD formation is required to match the observed orbital properties of WD--LC candidates in DR3. We further show that Gaia BH3-like binaries can form through standard isolated binary evolution without invoking any additional modelling assumptions, whereas reproducing Gaia BH1 and BH2 remains challenging within this framework. Looking ahead to the EOM, we predict detection of $\sim30$--$300$ BH--LCs, $\sim1500$--$5000$ NS--LCs, and $\sim10^5$--$10^6$ WD--LC binaries, primarily due to the significantly longer observational baseline.
\end{abstract}

\section{Introduction} 
\label{sec:intro}

\gaia, with its third data release (DR3), makes a significant leap in both the quality of data and the range of parameters available for each observed source. The primary objective of the mission is to study the structure, chemical composition, and star formation history of the Milky Way (MW). By mapping the distribution of positions, kinematics, and properties of stellar populations, \gaia\ is expected to provide an unprecedented map of the MW \citep[][]{Gaia_2000, Gaia_2016, Brown_2025, Perryman_2025}.

The first two data releases (DR) provided parallax and proper motion measurements for about a billion sources \citep[][]{Gaia_Collaboration_2016, Gaia_Collaboration_2018}. With the release of early DR3 and, more recently, DR3, there is an expansion in both the variety of sources and the diversity of their observed properties \citep[][]{Gaia_Collaboration_2021, Gaia_Collaboration_2022}. The DR3 catalog, in particular, features low-resolution blue photometry ($30\le R_{BP}\le100$), red photometry spectra ($60\le R_{RP} \le 90$) \citep[][]{DeAngeli_2023}, higher resolution ($R\sim11,500$) radial velocity spectrometer (RVS) spectra for sources with $G_{\rm{RVS}}\le14$ and their derivative information including, mean radial velocities \cite[][]{Katz_2023}, line broadening, and chemical composition  \citep[][]{Fremat_2023}. DR3 also includes photometric time-series data for sources in the Andromeda galaxy \citep[][]{Evans_2023}. DR3 not only improves the parameter estimates of the observed sources but also includes classifications of non-single stars \citep[\nss;][]{Arenou_2023}, eclipsing binaries \citep[][]{Mowlani_2023}, variable sources \citep[][]{Eyer_2023, Rimoldini_2023}, quasars and galaxies \citep[][]{Ducourant_2023}, and solar system objects \citep[][]{Tanga_2023}.

While previous data releases (e.g., DR1, DR2, and EDR3) focus primarily on single stars, DR3’s \nss\ catalog includes unresolved binary sources identified using astrometric, spectroscopic, and photometric techniques, based on the two-body orbital solution. The binary catalog provides inferred parameters such as the orbital period ($\porb$), eccentricity ($\ecc$), mass ratio, and component masses \citep[][]{Arenou_2023}. The \nss\ catalog features various types of binary star candidates, including ellipsoidal variables, cataclysmic variables, compact object–luminous companion (CO–LC) binaries, and stars with sub-stellar companions. However, given the quality of the orbital solutions, only a small fraction of the actual binary data is made public so far \citep[][]{Halbwachs_2023}.

With the release of \gaia's DR3 catalog, several important discoveries are made, including exoplanets, quasars, galaxies, variable sources, and COs \citep[][]{Bailer-Jones_2023, Delchambre_2023, Holl_2023a, Evans_2023, Eyer_2023, Wyrzykowski_2023}. Among these, the most noteworthy is the detection of two dormant black hole–luminous companion (\bhlc) candidates binaries from Gaia DR3 \citep[][]{El-badry2022e, El-badry_2023, Chakrabarti_2023, Tanikawa_2023}. Additionally, the catalog features approximately 3,200 white dwarf–luminous companion (WD–LC) candidates and a few dozen detached neutron star–luminous companion (NS–LC) candidates \citep[][]{Andrews_2022a, Shahaf_2022, Shahaf_2024, El-badry_2024a, El-badry_2024b}.

Several detections of detached \colc\ binary candidates and triple systems are reported over the past decade, primarily via spectroscopic observations, e.g., MWC\,656 \citep[][]{Casares_2014}, LB-1 \citep[][]{Liu2019}, 2MASS J05215658+4359220 \citep[][]{Thompson2019, Thompson_2020}, HR 6819 \citep[][]{Rivinius_2020}, HD96670 \citep[][]{Gomez_2021, WangH_2022}, V723 Mon \citep[][]{Jayasinghe2021}, NGC 1850 BH1 \citep[][]{Saracino2021},  NGC 2004\#115 \citep[][]{Lennon2021}, LTD064402+245919 \citep[][]{Yang_2021}, V1315 \citep[][]{Zak_2023}, ALS 8814 \citep[][]{An_2025} and BE Lyncis \citep[][]{Niu_2026}. However, follow-up observations of these candidates do not provide sufficient evidence for the presence of a CO, leaving their characterization debatable \citep[][]{Abdul-Masih_2020, Bodensteiner_2020a, El-badry_2020, El-Badry_2021, El-Badry2022a, El-Badry2022b, El-Badry2022c, El-badry_2025, Gies_2020, Heuvel_2020, Irrgang_2020, Mazeh_2020, Safarzadeh_2020, Shenar_2020, Simon_2020, Romagnolo2021, Guo_2022, Frost_2022, Rivinius_2022, Janssens_2023, Bianchi_2024, Kochanek_2025, Naze_2025, Nagarajan_2026}. Notably, the only confirmed detections—AS 386 \citep[][]{Khokhlov_2018} in the Galactic field and two detached BHs in the MW globular cluster NGC 3201 \citep[][]{Giesers2018, Giesers2019}—are reported prior to the release of the \gaia's \nss\ catalog. More recently, wide-area surveys have uncovered additional candidates, including VFTS 243 \citep[][]{Shenar_2022a}, VFTS 514 and 779 \citep[][]{Shenar_2022b}, HD 130298 \citep[][]{Mahy_2022}, 2XMM J125556.57+565846.4 \citep[][]{Mazeh_2022}, LAMOST J112306.9 + 400736 \citep[][]{Yi_2022},  LAMOST J235456.73+335625.9 \citep[][]{Zheng_2023}, 2MASS J06163552+2319094 \citep[][]{Yuan_2022}, 2MASS J15274848+3536572 \citep[][]{Lin_2023}, 56 UMa \citep[][]{Escorza_2023}, and  G3425 \citep[][]{WangS_2024}, VFTS 812 \citep[][]{Deshmukh_2026}, using spectroscopy and photometry.

The discovery of detached \colc\ binary candidates in the DR3 data identified via astrometry demonstrates \gaia's ability to probe these systems, as predicted by numerous theoretical studies \citep[e.g.,][]{Barstow2014, Kawanaka_2016,  Breivik_2017, Mashian2017, Kinugawa2018, Streseman2018, Yalinewich2018, Yamaguchi2018,  Shao_2019, Shikauchi2020, Shikauchi_2022, Shikauchi_2023, Wiktorowicz2020, Chawla2022, Janssens_2022, Wang_2022}. It is now evident that a large population of detached \colc\ binaries awaits discovery in future data releases \citep[][]{Brown_2025, El-badry_2024e, Perryman_2025}. However, it is not straightforward to compare the theoretical population studies with the observed because the theoretical studies focused on the end-of-mission (EOM) detectable populations. These recent findings thus motivate a DR3-specific analysis of the modeled populations of detached \colc\ binaries. Moreover, we expand our populations to include {\em all} CO-LC binaries without restricting to BH-LC binaries only \citep[e.g., in][]{Chawla2022,Chawla_2023}.  

We present the intrinsic detached CO-LC populations expected to be present in the MW. In addition, we identify the populations that would be detectable by \gaia's astrometry by the EOM. Moreover, we present the populations that are specifically identifiable under \gaia's DR3-specific observational constraints. In \autoref{sec:numerical setup}, we describe our numerical framework for generating the CO–LC populations and constructing synthetic MW models. Section \ref{sec:detection-criteria} outlines the criteria for resolving a CO–LC binary with \gaia, accounting for astrometric detectability, interstellar extinction, and reddening. In \autoref{sec:Results}, we present the key properties of the intrinsic and \gaia-resolvable CO–LC populations, both for EOM and DR3. Section \ref{sec:comparison with CO--LC candidates} compares the properties of the \colc\ candidate binaries identified using \gaia's DR3 with those of our model populations. We summarize our key findings and conclude in \autoref{sec:conclusion}.

\section{Numerical Setup} \label{sec:numerical setup}

We use \cosmic\ \citep[][]{Breivik_2020, Breivik_2021}, a publicly available rapid binary population synthesis suite, to model highly realistic populations of \bhlc, \nslc, and \wdlc\ binaries in the MW. \cosmic\ is based on the stellar evolution fits from \sse\ \citep[][]{Hurley2000} and binary evolution algorithm \bse\ \citep[][]{Hurley2002}. The current version of \cosmic\ incorporates various modifications from \sse/\bse\ for the stellar evolution models and to treat binary interaction processes; for a detailed review, we encourage readers to see \citet[][]{Breivik_2020}.

Using \cosmic, we model the present-day \colc\ binary populations by evolving an initial Zero Age Main Sequence (ZAMS) binary population. We generate the ZAMS binary stellar and orbital parameters from various observationally motivated probability distribution functions by adopting the `independent' sampling scheme of \cosmic\ which assumes no correlations between stellar parameters (e.g., primary mass, mass ratio) and orbital parameters (e.g., semi-major axis, eccentricity) at birth. We sample the primary mass $\mprim$ between $0.08$ and $150\,\msun$ from the initial stellar mass function given in \citet{Kroupa2001}. We assign the secondary masses $\msec$ such that $\msec/\msun\geq0.08$ and the mass ratio $q\equiv\msec/\mprim$ distribution is uniform \citep[][]{Mazeh1992, Goldberg1994, Kobulnicky_2007}. We assume an initially thermal eccentricity distribution \citep[][]{Jeans_1919, Ambartsumian_1937, Heggie1975}. The initial semi-major axis ($\sma$) is sampled from a log-uniform distribution \citep[][]{Abt1983} with an upper bound of $10^{5}\rm{R_{\odot}}$ and a lower bound set by $\rpri\le\rRoche/2$ \citep[][]{Dominik_2012}, where $\rpri$ and $\rRoche$ denote the radius and Roche radius of the primary. 

We use a metallicity-dependent star formation rate based on the final snapshot of the synthetically modeled MW-like galaxy {\bf m12i} in the Latte suite of the Feedback In Realistic Environments 2 (FIRE-2)\footnote{http://fire.northwestern.edu} simulation suite \citep{Hopkins2014, Hopkins2018, Wetzel2016}. We further use the \textsf{ananke} framework to assign three-dimensional Galactic positions for each binary by resampling from the smoothed density distribution of the star particles \citep[\autoref{sec:Synthetic MW};][]{Sanderson2020}. This approach preserves the complex correlations between age, metallicity ($\metal$), and the spatial distribution of the star particles in {\bf m12i}. This simulation does not account for the potential impact of natal kicks which may alter the final positions of the star particles that are used in our study \citep{Wagg_2025}. However, we note that \citet[][]{Wagg_2025} focus exclusively on the population of massive stars which originate in disrupted binaries. In contrast, we are only concerned with detached \colc\ binaries that survive the natal kicks and hence should have much lower systemic velocities compared to the systems studied by \citet{Wagg_2025}. Finally, we note that \bse\ restricts the metallicity values to the range $10^{-4}\le Z\le 0.03$; therefore, \cosmic\ assigns these limiting values to binaries when the sampled metallicity falls outside this range.

Various binary interaction processes, for example, mass transfer (MT), common envelope (CE) evolution, and tides, play a crucial role in shaping the present-day properties of the \colc\ binary populations. We use the adiabatic response of the donor star to determine whether MT through Roche lobe overflow (RLOF) is stable or leads to CE \citep[][]{Soberman1997}. We assume critical mass ratios as a function of donor type from \citet[][]{Belczynski2008}. If a binary enters a CE, we estimate the post-CE state using the ``energy formalism" \citep[][]{Livio1984, Webbink_1984} which uses two parameters, $\alpha$, the efficiency with which the available energy transfers to the envelope for ejection, and $\lambda$, a measure of the binding energy of the envelope relative to the gravitation potential energy of the binary orbit, to predict the final orbital eccentricity and orbital period. We adopt the standard value $\alpha=1$ \citep[][]{Dominik_2012}, whereas $\lambda$ depends on the present-day evolutionary stage of the star as described in \citet[][]{Claeys2014}.

\subsection{Compact-Object Formation: White dwarfs}
\label{sec:WD-form}
Three types of WDs can be produced in our models depending on metallicity, ZAMS mass, and binary interactions. According to stellar evolution theory, low-mass 
($0.08\lesssim\mzams/\msun\lesssim0.8$) stars leave behind a He-WD after the He core contracts until supported by electron degeneracy pressure. Low-mass stars 
($0.8\lesssim\mzams/\msun\lesssim2$) ignite He via the so-called He flash, while intermediate-mass stars ($2\lesssim\mzams/\msun\lesssim8$) burn He in their cores; in both cases a carbon-oxygen core forms which is eventually supported by electron degeneracy pressure, leaving behind a CO-WD. Higher-mass stars ($8\lesssim\mzams/\msun\lesssim10$) ignite both He and carbon in their cores, and their fate is determined by the mass of the resulting neon-oxygen-magnesium core. If this core mass exceeds $\mch\approx1.38\,\msun$ \citep{Chandrasekhar_1931}, electron capture initiates collapse to form a low-mass NS 
\citep[e.g.,][]{Miyaji_1980, Nomoto_1984, Nomoto_1987}; otherwise, a oxygen-neon (ONe)-WD remains.\footnote{The $\mzams$ ranges are indicative and the nature of the remnant depends on the pre-collapse core mass. As a result, metallicity-dependent stellar winds and binary interactions can significantly blur these boundaries \citep[][]{Mink_2009, Langer_2012, Agrawal_2022, Eldridge_2022}.}

In \cosmic, the WD type is determined by the nuclear-burning stage and core mass at the time the envelope is lost, rather than by the ZAMS mass alone \citep{Hurley2000}. If the envelope is removed early---while the star is still in the Hertzsprung gap or on the giant branch---the remnant is a He WD, provided the degenerate core lies below $M_\mathrm{HeF}$ (the maximum initial mass for which helium ignites degenerately in a flash); otherwise the star is left as a naked helium star whose degree of evolution depends on the stage at which the envelope was stripped \citep{Hurley2000}. Such early envelope loss is driven primarily by binary interactions. He-WDs can therefore only be produced within the \cosmic/\bse\ framework through binary interaction. CO and ONe-WDs, in contrast, form when the envelope is lost later, before the core mass reaches $M_\mathrm{c,SN} \approx \mch$ on the thermally pulsating asymptotic giant branch (TPAGB), and their classification is set by the hydrogen-exhausted core mass at the base of the AGB ($M_\mathrm{c,BAGB}$). The remnant is a CO WD for $M_\mathrm{c,BAGB} < 1.6\,\msun$. If the CO core mass $M_\mathrm{c,CO}$ exceeds $1.08\,\msun$ carbon ignites off-centre under semi-degenerate conditions producing a degenerate ONe core\citep{Pols1998, Nomoto_1984}. Depending on the metallicity, this translates to $1.6\,\msun \leq M_\mathrm{c,BAGB} \leq 2.25\,\msun$. Above the upper limit, $M_\mathrm{c,BAGB} > 2.25\,\msun$, the core no longer ends as a WD: \cosmic/BSE instead forms NSs and, at the highest masses, BHs (see \autoref{sec:SN-phy}). While the limiting core masses $M_\mathrm{c,BAGB} = 1.6$ and $2.25\,\msun$ are essentially independent of metallicity, the corresponding ZAMS masses are dependent on metallicity \citep{Pols1998}. These two limits mark, respectively, the minimum initial mass undergoing non-degenerate carbon ignition and the minimum initial mass that avoids electron capture on Ne and Mg in the core.

The other difference between CO and ONe WDs in \cosmic\ is in their response to mass accretion \citep{Hurley2000}. WD accretion is implemented following \citet{Hurley2002}, where the fraction of transferred mass accreted onto the WD is determined by comparing the mass transfer timescale $\tau_{\dot{M}} = M_2/\dot{M}_t$ with the Kelvin-Helmholtz timescale $\tau_{\rm KH2}$ of the accretor. If $\tau_{\dot{M}} \ll \tau_{\rm KH2}$, the system enters a common-envelope state; if $\tau_{\dot{M}} \gg \tau_{\rm KH2}$, all transferred mass is accreted. In the intermediate regime, only a fraction $1-f$ of the transferred mass is accreted, where:
\begin{equation}
    1 - f = \min\left(10\frac{\tau_{\dot{M}}}{\tau_{\rm KH2}},\ 1\right).
\end{equation}
Mass accretion is also restricted by the Eddington limit \citep{Cameron1967}:
\begin{equation}
    \dot{M}_{\rm Edd} = \frac{4\pi c R_2}{\kappa} = 2.08\times10^{-3}(1+X)^{-1}
    \frac{R_2}{R_\odot}\ M_\odot\ {\rm yr}^{-1},
\end{equation}
where $X$ is the hydrogen mass fraction, $\kappa$ is the opacity, and $R_2$ is the WD radius; this limit is used in place of the above prescription for degenerate accretors. 

In our simulations, a WD can grow by accreting material directly onto its degenerate core. If accretion drives the WD mass above $\mch$, a CO WD is destroyed in a thermonuclear (Type Ia) explosion leaving no remnant \citep[][]{Livne_1995}, whereas an ONe WD undergoes electron capture on $^{24}$Mg nuclei forming a NS via accretion-induced collapse \citep[AIC;][]{Nomoto_1991, vanParadijs1997}. We assume that WDs receive no natal kicks in our fiducial models; however, we explore the effects of moderate natal kicks in \autoref{sec:WD_DR3}.

\subsection{Compact-Object Formation: Neutron stars and black holes}
\label{sec:SN-phy}

Stars with ZAMS mass $\mzams/\msun\gtrsim10$ are massive enough to ignite the elements beyond carbon, ultimately forming an iron core. At collapse, the final structure of these stars is composed of an iron core surrounded by shells of various elements produced as a by-product of nuclear burning at different stages of a star's life. When such a star explodes, it leaves behind a NS or a BH \citep[][]{Fryer_1999, Fryer_2001, Heger_2003}.

We adopt two widely used explosion mechanisms for NS or BH formation through core-collapse supernovae (CCSNe), `rapid' and `delayed' \citep[][]{Fryer2012}. These mechanisms differ in the explosion time after the core bounce set by the timescales on which neutrino-driven instabilities are allowed to grow \citep[][]{Colgate_1966, Boccioli_2025}. In this work, we adopt $3\,\msun$ as the upper mass limit for NSs, so that any compact object with $\mco/\msun > 3$ is classified as a BH \citep[][]{Chawla2022, Chawla_2023}. This cut sets the floor of the BH mass distribution, and whether remnants populate the region just above it depends on the explosion mechanism. In the rapid prescription, the short explosion time leaves a dearth of births between this $3\,\msun$ limit and $5\,\msun$--creating a `mass-gap' in the BH mass function at birth \citep[][]{Bailyn_1998, Ozel_2010, Farr_2011, Kreidberg_2012, Belczynski_2012}. Even in the \rapid\ model, however, this gap need not remain empty: a NS that accretes beyond the maximum NS mass can undergo AIC to form a low-mass BH \citep[][]{Shibata_2003, Perna_2021}, populating the mass gap---a consequence of binary interaction rather than standard core-collapse supernovae. The delayed prescription, in contrast, fills this region with continuous remnant births and leaves no mass gap. Finally, NSs may also form via ECSN, through the same electron-capture collapse of a degenerate ONe core (see \autoref{sec:WD-form}). The relevant parameter is the He-core mass range that fails to reach iron-core formation, which we adopt to be $1.4$--$2.5\,\msun$ \citep[][]{Podsiadlowski_2004}.

Due to deviations from spherical symmetry during the collapse and subsequent explosion, the stellar remnant may experience a kick at the time of formation \citep[][]{Fryer_2001, Janka_2013}. For NSs, we adopt a bimodal Maxwellian distribution with $\sigma=265$ and $20\,\kmps$ for CCSN and ECSN, respectively \citep[][]{Hobbs2005, Podsiadlowski_2004}. For BHs, we assume a fallback-modulated kick magnitude, where the initial kicks are sampled from a Maxwellian distribution with $\sigma=265\,\kmps$ \citep[][]{Belczynski_2012, Fryer2012, Nuchvanichakul_2025}.

\subsection{Synthetic Milky Way population \label{sec:Synthetic MW}}

\cosmic\ iteratively simulates binaries until the distributions of observable parameters including $\mco$, $\mlc$, $\porb$, and $\ecc$ converges to a high degree of resolution. We quantify the convergence of the distributions using a `$\match$' parameter inspired by the matched filtering technique \citep[][]{Chatziioannou2017, Breivik_2020}. For this study we have adopted $\log_{10}(1-\match)=-5$. The matched-filtering technique enables \cosmic\ to evolve only a fraction of the MW mass to generate a population of \colc\ binaries, referred to as the intrinsic population, which accurately captures the parameter space of the full \colc\ binary population in the MW. To estimate the total number of \colc\ binaries in the MW, we scale the intrinsic \colc\ population by a factor equal to the ratio of the MW mass ($M_{\rm{\bf{m12i}}}$ as specified by {\bf m12i}) to the total simulated mass ($M_{\rm{sim}}$ including both single and binary stars determined based on the adopted binary fraction) which created the converged intrinsic \colc\ population. The number of \colc\ binaries in the MW ($\ncolcmw$) can then be estimated from the number of simulated \colc\ binaries ($\ncolcsim$) as-

\begin{equation}
   \ncolcmw = \ncolcsim \frac{M_{\rm{\bf m12i}}}{M_{\rm{sim}}}.
\end{equation}

We sample $N_{\colc,\rm{MW}}$ binaries with replacement from the intrinsic population of \colc\ binaries assigning all the key stellar and orbital properties. Each binary is also given a unique position in the MW galaxy based on the age and metallicity of the ZAMS progenitor. This preserves the complex correlations between the position, age, and metallicity of the different star particles in ${\bf m12i}$. Every binary is also given a random orientation with respect to the line of sight. Based on the orientation of the binary, we calculate its sky-projected orbit using the Thiele-Innes elements \citep[see][for a detailed description]{Chawla2022}. Finally, we create multiple Galactic realisations of the \colc\ binary populations in order to incorporate the effects of statistical fluctuations due to distribution of the binaries in the MW and their orbital orientations. We create $200$, $100$ and $3$ Galactic realisations for the \bhlc, \nslc\, and \wdlc\ binaries, respectively. 

We evaluate the \gaia-relevant magnitudes of the LCs from the bolometric luminosities and effective temperatures given by \cosmic\ using the {\tt isochrones} package \citep[][]{Isochrones_2015}. We correct these magnitudes to incorporate the effect of reddening and extinction due to interstellar dust using the three-dimensional `combined19' dust map included in the $\mwdust$ package \citep[][]{Drimmel2003, Marshall2006, Green2019, Bovy2016} given the distance and Galactic location of each binary.

\section{Detectibility by \gaia} 
\label{sec:detection-criteria}
In this section, we describe the detectability criteria that we adopted to characterize a binary as resolvable by \gaia, based on magnitude $G$ and the astrometric motion of the LC due to its orbit around the CO. We separately consider detectability of binaries resolvable by EOM and within the more stringent restrictions of DR3. 

Note that the \gaia\ selection function used in this study is approximate, as it compares the sky-projected angular size with the \gaia\ pointing error. It is, however, computationally inexpensive and thus suitable for our large-scale simulations. More sophisticated \gaia\ selection functions, which incorporate additional complexities such as the scanning law, crowding, and photometric systematics, have been developed following the release of DR3 \citep[][]{El-badry_2024d, Lam_2025}. While our simplified approach does not capture these effects in detail, they are unlikely to significantly alter our population-level conclusions. Moreover, our predicted detection rates are consistent with theoretical studies employing more refined approaches \citep[][]{Nagarajan_2025a, Yamaguchi_2025} as well as with the currently observed census of \colc\ binaries in the DR3 data \citep[][]{Shahaf_2022, Shahaf_2024, El-badry_2024a, El-badry_2024b}.

\subsection{End-of-mission resolvable population}
\label{sec:detection-criteria-dr5}
We employ three important observational cuts while defining the resolvability of the detached \colc\ binaries by \gaia's astrometry: 
\begin{enumerate}
    \item $\porb/\yr\le10$,
    \item $G \le20$,
    \item $\alpha\ge k\sigma_{G}$,
\end{enumerate}
where $\alpha$ is the sky-projected angular size of the LC orbit, $\sigma_{G}$ is \gaia's pointing precision as a function of $G$ \citep[][]{Lindegren2018}. We create two different sets, $k=1$ ($3$) for optimistic (pessimistic) case. These observational cuts are identical to those adopted in \citet{Chawla2022} where a more detailed description is given. We refer to these sets as optimistic and pessimistic `\gaia-resolvable' \colc\ binaries.     

\subsection{DR3 selection cuts}
\label{sec:detection-criteria-dr3}
\gaia's \nss\ catalog includes sources that do not have optimal single-star solutions. The input source list used to create the \nss\ catalog was selected from single-star solutions with renormalized unit weight error $\rm{RUWE}\geq1.4$ \citep[][]{Belokurov_2020, Penoyre_2020, Penoyre_2022, Penoyre_2022a, Gandhi_2021, Andrew_2022, Arenou_2023}, indicating that these sources might either be a binary star or have a problematic astrometric single-star solution. This input list gets further filtered using additional cuts on available parameters, including $\texttt{ipd\_frac\_multi\_peak} \leq 2$, $\texttt{ipd\_gof\_harmonic\_amplitude} < 0.1$, $\texttt{visibility\_periods\_used} > 11$, and $|C^{*}| < 1.645\sigma_{C^{*}}$ \citep[][]{Arenou_2023, Halbwachs_2023}, ultimately selecting sources with $G \leq 19$. These cuts were designed to reduce contamination from partially resolved pairs and sources with nearby bright neighbours, whose scan-angle-dependent astrometric signals can mimic binary orbital motion \citep[][]{Holl_2023a}.

Despite the stringent selection criteria, the output dataset still contained a significant fraction of spurious binary solutions \citep[][]{Arenou_2023}. These spurious solutions arise primarily from the \gaia\ nominal scanning law \citep[][]{Holl_2023a}, which has the following key properties relevant for spurious solution generation:
\begin{enumerate}
    \item The spacecraft's scan angle at any given sky position varies with a dominant yearly modulation and a $\sim$63\,days spin-axis precession period, causing scan angles to cluster near the ecliptic plane due to a geometric circle of avoidance.
    \item For non-point-like sources — such as unresolved or partially resolved pairs — the image parameter determination procedure fits a single point-source model to a blended profile, introducing a position bias that varies sinusoidally with scan angle.
    \item The resulting spurious orbital solutions cluster at specific periods satisfying $365.25/P = m \times 5.8 + n$, where $m$ and $n$ are small integers and 5.8\,cycles\,yr$^{-1}$ is the precession frequency of the spin axis \citep[][]{Holl_2023a}.
\end{enumerate}
A majority of these spurious solutions, particularly those with periods below $\sim$100\,days, resulted in unreasonably large mass functions \citep[][]{Arenou_2023, Halbwachs_2023}. This occurs because the spurious semi-major axis $a_0$ is set by the amplitude of the scan-angle bias rather than by Kepler's third law — it is therefore decoupled from the fitted period, and dividing an arbitrary $a_0$ by a small $P^2$ in the mass function formula produces an unphysically large $f_{\mathcal{M}}$.

To filter out these spurious solutions, the DR3 pipeline adopts a period-dependent parallax significance criterion:
\begin{equation}
    \frac{\Pi}{\sigma_{\Pi}} \geq 
    \frac{20{,}000\,\rm{days}}{\porb},
    \label{eq:dr3-parallax-cut}
\end{equation}
%i
which was found to effectively discard the majority of spurious solutions concentrated at short periods \citep[][]{Halbwachs_2023}, while preserving real binary solutions around nearby stars with well-measured parallax. 

We identify the synthetic \colc\ populations detectable in DR3 by adopting selection cuts in addition to those adopted for identifying the EOM \colc\ populations (\autoref{sec:detection-criteria-dr5}) based on the selection cuts relevant specifically for DR3.
\begin{enumerate}\setcounter{enumi}{3}
    \item $G \leq 19$,
    \item $\porb/\yr \leq 3$,
    \item $\Pi/\sigma_{\Pi} \geq 
    \frac{20{,}000\,\rm{days}}{\porb}$,
\end{enumerate}
where $\Pi$ and $\sigma_{\Pi}$ denote parallax and parallax measurement precision, respectively. We refer to the set of \colc\ binaries satisfying these three additional conditions as `DR3-resolvable.' Note that in the DR3-resolvable population, after applying the additional cuts described above, the optimistic and pessimistic datasets merge. This is because the binaries satisfying DR3-resolvable populations are wide enough that the entire optimistic dataset also satisfies $\alpha \geq 3\sigma_G$.
\section{Results \label{sec:Results}}
In this section, we describe the properties of the intrinsic CO-LC populations. We also discuss the overall expected detection rate of the \colc\ binary population based on \gaia's EOM observations, as well as the expected detections from the DR3 data release. We further discuss the possibilities of forming the \gaia\ observed \colc\ candidates, considering both the isolated and dynamical channels. Finally, we provide comparisons of the stellar and orbital characteristics of the \gaia\ observed \colc\ candidates with our modeled population. The sizes of the \gaia-detectable CO--LC populations for each set of binary models and observational selection cuts are summarized in \autoref{tab:detection}.

\begin{deluxetable*}{c|c|c|c|c|c|c|c}
\tablecolumns{6}
\tabletypesize{\normalsize}
\tablecaption{CO-LC binaries in the Milky Way} 
\tablehead{
 \colhead{CO Type}\vline & \colhead{SN Model}\vline &
 \colhead{LC Type}\vline & \colhead{Intrinsic ($\porb/\yr\le 10$)}\vline &
 \multicolumn{4}{c}{\gaia\ Resolvable} \\
 \cline{1-8}
  \colhead{}\vline & \colhead{}\vline &
 \colhead{}\vline & \colhead{}\vline &
 \multicolumn{2}{c}{Optimistic}\vline & \multicolumn{2}{c}{Pessimistic}\\
 \cline{1-8}
 \colhead{}\vline & \colhead{}\vline &
 \colhead{}\vline & \colhead{}\vline &
 \colhead{EOM }\vline & \colhead{DR3}\vline & \colhead{EOM }\vline & \colhead{DR3}
 }
\startdata
  & & MS & $8658^{+113}_{-123}$ & $255^{+19}_{-26}$ & $0$ &
  $139^{+12}_{-18}$ & $0$\\
  & \rapid & PMS & $529^{+26}_{-27}$ & $36^{+9}_{-7}$ & $0$ &
  $14^{+5}_{-4}$ & $0$\\
   BH & & Total & $9184^{+120}_{-133}$ & $292^{+21}_{-26}$ & $0$ &
   $152^{+14}_{-17}$ & $0$\\
%   %
  & & MS & $6653^{+106}_{-104}$ &
  $68^{+11}_{-9}$ & $0$ &
  $26^{+7}_{-6}$ & $0$ \\
  & \delayed\ & PMS & $443^{+26}_{-28}$ & $27^{+6}_{-7}$ & $0$ &
  $9^{+3}_{-4}$ & $0$\\
  & & Total & $7091^{+113}_{-100}$ & $95^{+12}_{-12}$ & $0$ &
  $36^{+7}_{-8}$ & $0$ \\  
%  %
    \hline
%  %
  & & MS & $98,199^{+245}_{-199}$ & $4553^{+73}_{-87}$ & $27^{+7}_{-6}$ &
  $1422^{+51}_{-46}$ & $27^{+7}_{-6}$\\
  & \rapid & PMS & $19,465^{+126}_{-110}$ & $290^{+27}_{-25}$ & $2^{+3}_{-1}$ &
  $105^{+12}_{-16}$ & $2^{+3}_{-1}$\\
  NS & & Total & $117,700^{+230}_{-247}$ & $4843^{+71}_{-84}$ & $29^{+9}_{-5}$ &
  $1525^{+49}_{-47}$ & $29^{+9}_{-5}$ \\
%   %
   % \hline
%   %
  & & MS & $102,101^{+224}_{-272}$ & $4695^{+78}_{-84}$ & $14^{+5}_{-6}$ & $1519^{+51}_{-34}$ & $14^{+5}_{-6}$ \\
  & \delayed\ & PMS & $20,100^{+97}_{-100}$ & $334^{+21}_{-25}$ & $0$ &
  $88^{+17}_{-9}$ & $0$  \\
  & & Total & $122,170^{+269}_{-242}$ & $5021^{+89}_{-74}$ & $14^{+5}_{-6}$ &
  $1612^{+52}_{-48}$ & $14^{+5}_{-6}$\\
  %  %
    \hline
%  %
  & & MS & $586,354,452$ & 
  $1,383,727^{+134}_{-892}$ & $3256^{+24}_{-16}$ &
  $405,246^{+1032}_{-141}$ & $3256^{+24}_{-16}$\\
  WD& & PMS & $14,943,842$ & $783,420^{+859}_{-453}$ & $932^{+63}_{-39}$ &
  $258,634^{+224}_{-483}$ & $932^{+63}_{-39}$\\
  & & Total & $601,298,294$ & $2,166,748^{+1178}_{-573}$ & $4188^{+87}_{-55}$ &
  $663,880^{+1256}_{-624}$ & $4188^{+87}_{-55}$\\
\enddata
\tablecomments{\colc\ numbers in the Milky Way predicted in our models. The numbers and errors denote the median and the spread between the $10$th and $90$th percentiles across the Milky-Way realisations ``Intrinsic" denotes the present-day detached population of CO-LC binaries with $\porb/\yr\le10$ in the Milky Way. ``Resolvable" denotes the subset of CO-LC binaries resolvable by \gaia's astrometry.
}
\label{tab:detection}
\end{deluxetable*}
\subsection{Intrinsic CO--LC population in the MW}
\label{sec:intrinsic-pop}

\begin{figure}
    \plotone{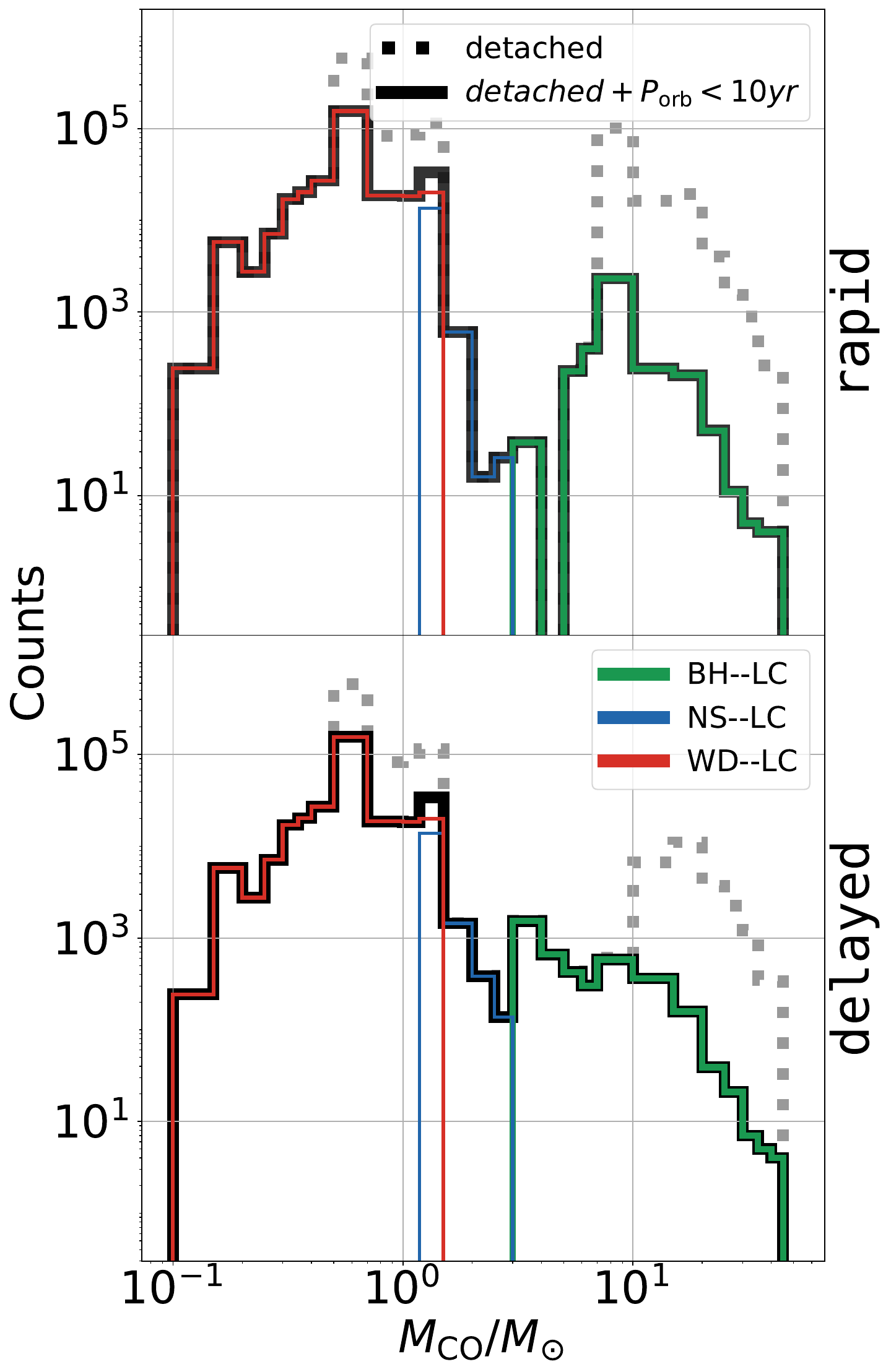}
    \caption{$\mco$ distribution of the present-day \colc\ binary populations in the MW. The dashed and solid curves represent only detached \colc\ and detached binaries with $\porb/\yr\le10$, respectively. The green, blue, and red curves represent \bhlc, \nslc, and \wdlc\ populations, respectively. }
    \label{fig:intrinsic-mco}
\end{figure}

The distributions of the stellar and orbital properties of the intrinsic \colc\ population provide important clues about the underlying binary interaction physics. Various binary interaction processes, for example, MT, CE evolution, tidal evolution, and SN, collectively shape the present-day characteristics of the \colc\ population. Analyzing the trends observed in the properties helps to understand how the parameter space of the intrinsic population is represented by the \gaia\ observed \colc\ population.

\begin{figure}
    \plotone{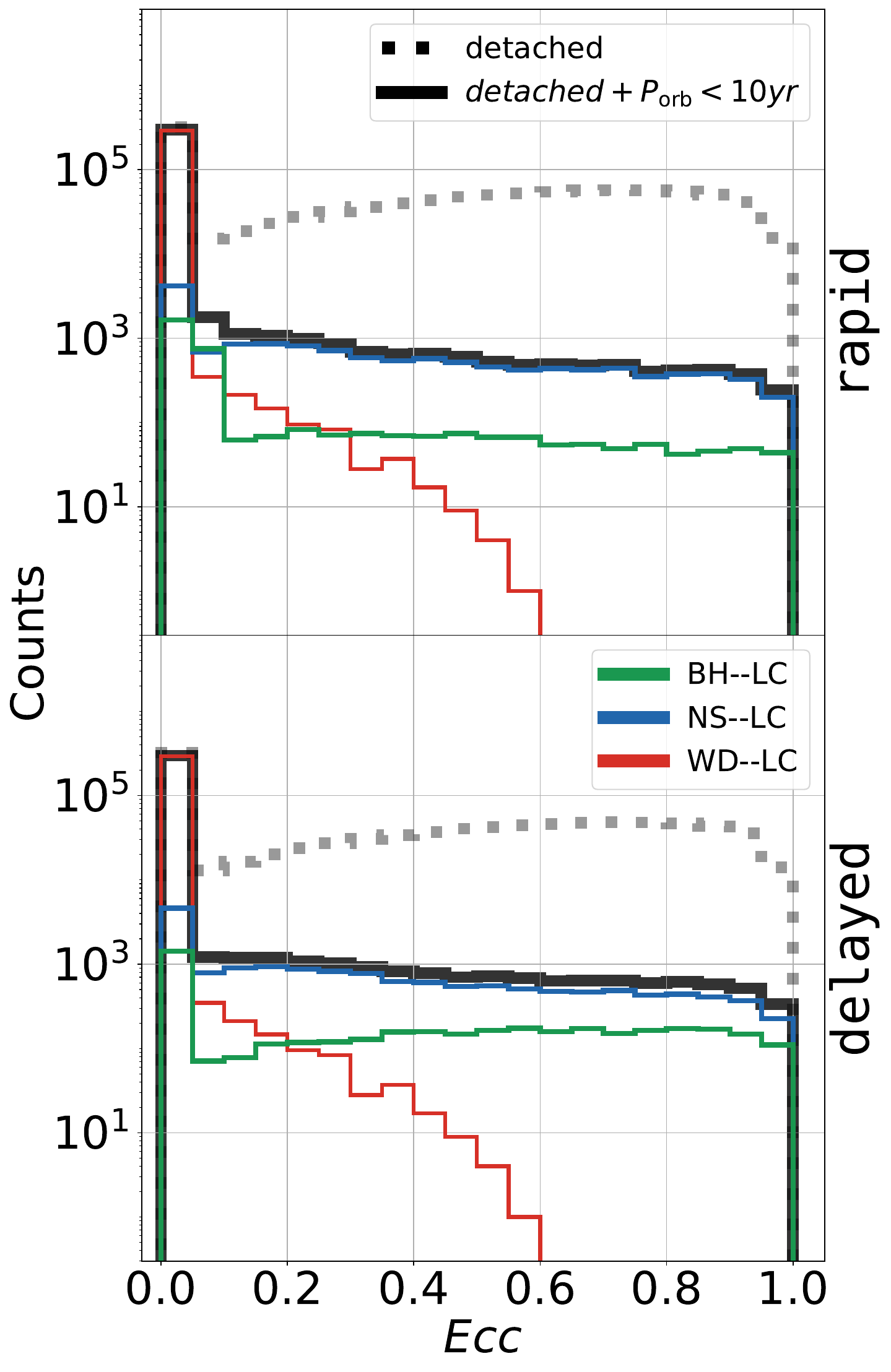}
    \caption{Same as \autoref{fig:intrinsic-mco} but for eccentricity distributions of the present-day \colc\ binary population in the MW.}
    \label{fig:intrinsic-ecc}
\end{figure}

\begin{figure}
    \plotone{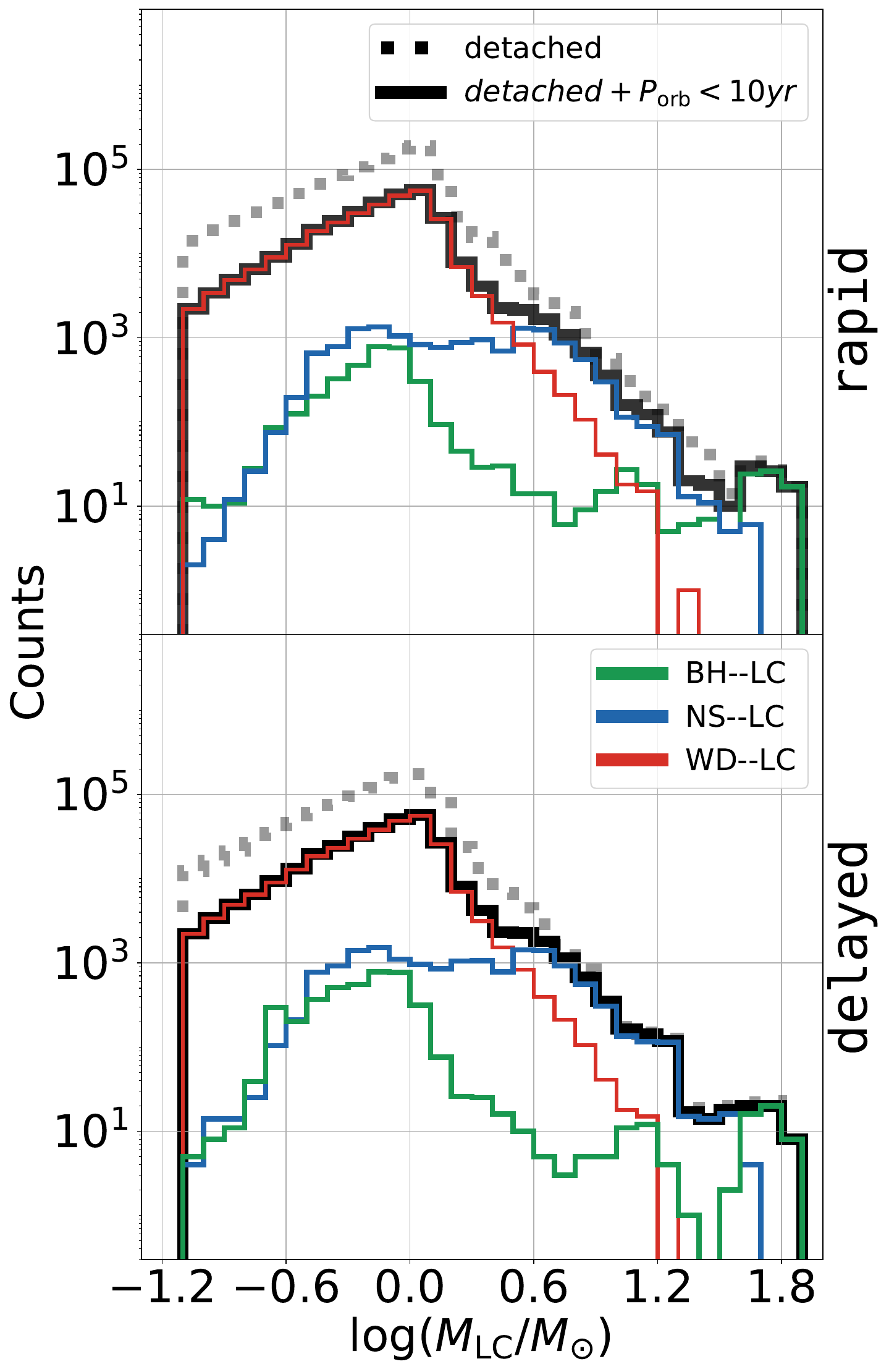}
    \caption{Same as \autoref{fig:intrinsic-mco} but for $\mlc$ distribution of the present-day \colc\ binary population in the MW.}
    \label{fig:intrinsic-mlc}
\end{figure}

\begin{figure}
    \plotone{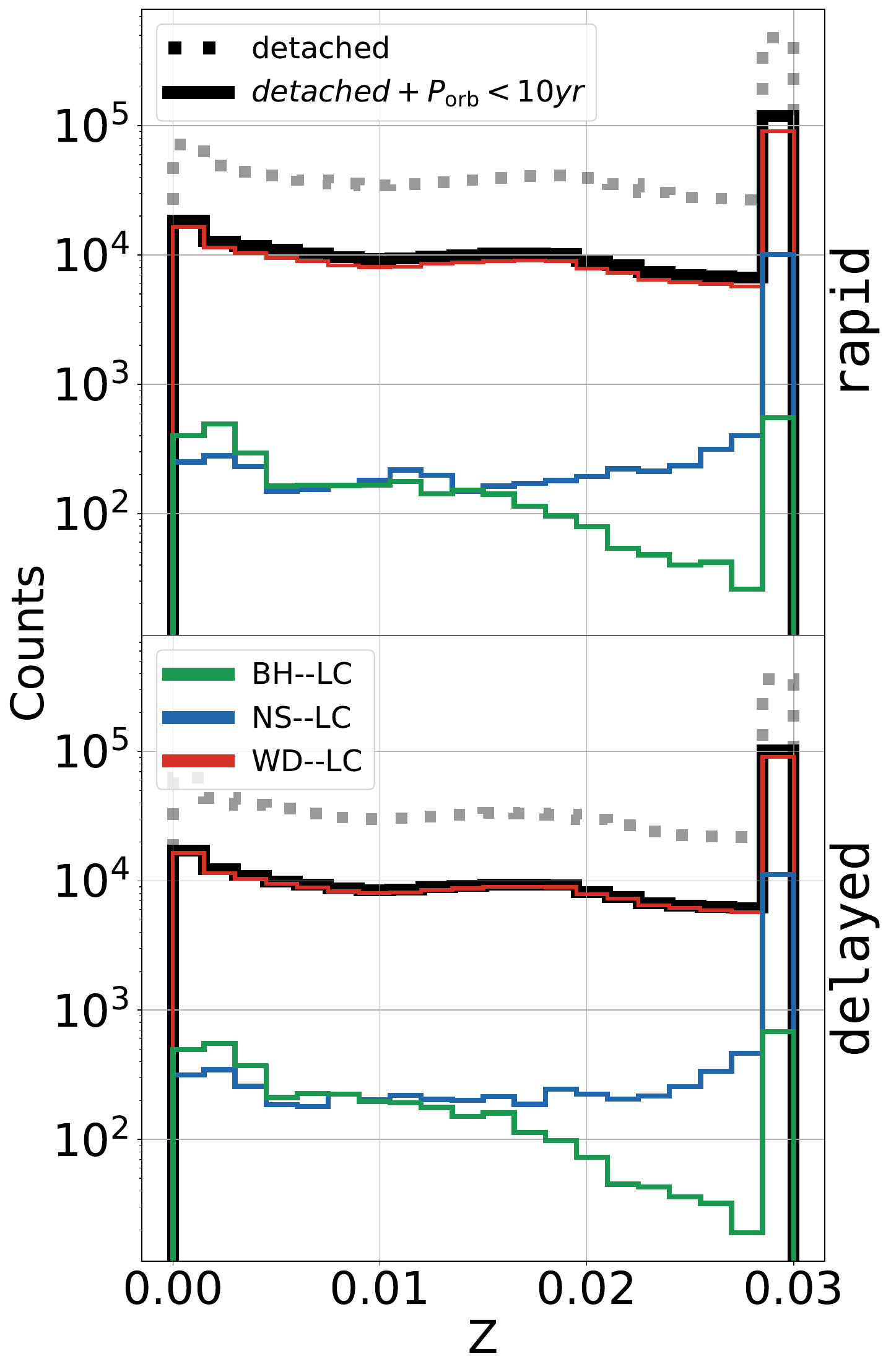}
    \caption{Same as \autoref{fig:intrinsic-mco} but for metallicity (Z) distribution of the present-day \colc\ binary populations in the MW.}
    \label{fig:intrinsic-met}
\end{figure}

\autoref{fig:intrinsic-mco}, \ref{fig:intrinsic-ecc}, \ref{fig:intrinsic-mlc}, and \ref{fig:intrinsic-met} show the distributions of $\mco$, $\ecc$, $\mlc$, and $\metal$ of the intrinsic \colc\ populations overall (dotted histograms) and those with $\porb/\yr\le10$ (solid histograms) for the adopted \rapid\ (top) and \delayed\ (bottom) SN prescriptions. The $\mco$ distribution (\autoref{fig:intrinsic-mco}) exhibits a wide range between $0.1-45\,\msun$ and clearly shows two segments. The \colc\ binaries with $\mco\le3\,\msun$ comprise of a combined population of \nslc\ and \wdlc\ binaries. The lower mass range $0.1\leq\mco/\msun\leq1.4$ primarily consists of the WDs along with a small fraction of low-mass NSs.The He, carbon-oxygen, and ONe WDs populate roughly within mass ranges of $0.1\le\mwd/\msun\le0.5$, $0.3\le\mwd/\msun\le1.1$, and $1.1\le\mwd/\msun\le\mch$, respectively, where the $\sim1.1\,\msun$ boundary between CO and ONe WDs corresponds to the off-centre carbon ignition threshold (see \autoref{sec:WD-form}).

The NS binaries contribute in the range $1.20\le\mns/\msun\le3$ based on our adopted lower and upper mass limits for NSs. For the \nslc\ binaries with $\porb/\yr\le10$ in our models, the contribution from the ECSN (CCSN) channel varies between $70$--$72\%$ ($27$--$29\%$) depending on the adopted SN prescription. The relatively higher contribution from the ECSN channel can be attributed to the significantly larger natal kicks the NSs receive during CCSN which often break the progenitor binaries. In addition, we find that $\lesssim 1\%$ of \nslc\ binaries with $\porb/\yr\le10$ form via AIC of a massive WD. While, the overall \nslc\ population contains a significantly more prominent contribution ($\sim25-26\%$) from massive NSs ($\mns/\msun\ge2.0$), these NSs are predominantly ($\sim93-99\%$) in actively accreting systems via RLOF in our models at present day. These systems are of course very interesting in general. However, since we focus on the {\em detached} \colc\ binaries exclusively in this study, these \nslc\ binaries do not satisfy our selection and hence do not contribute significantly in the distributions shown in \autoref{fig:intrinsic-mco}. The most prominent difference between the rapid and delayed models can be seen in the mass range $3\le\mco/\msun\le5$ as expected; while the rapid model creates a mass gap between NSs and BHs created via CCSN within this range, the \delayed\ model does not. BHs with $3\le\mbh/\msun\lesssim4$ for the \rapid\ population come from AIC of NSs. 

Eccentricity is another parameter that can differentiate between the two adopted SNe prescriptions (\autoref{fig:intrinsic-ecc}). While the overall population of \colc\ binaries can have significant eccentricities, the eccentricities for those with $\porb/\yr\le10$ remains low. This is because most of these binaries experience at least one MT episode which erases the initial eccentricity. The present-day eccentricities for these binaries are set solely by the adopted natal kicks and binary interactions post the formation of the CO. Although natal-kick physics remains uncertain and different kick prescriptions can lead to variations in post-SN eccentricities and $\porb$ of individual binaries, previous studies have shown that the overall shapes of the orbital parameter (e.g., $\porb$ and $Ecc$) distributions of the Gaia-detectable \bhlc\ and \nslc\ populations are not strongly affected by the choice of kick model \citep{Breivik_2017, Yalinewich2018}.

In absence of natal kicks in our fiducial model, the eccentricities of \wdlc\ binaries remains low and depend solely on the details of binary interaction physics including MT, CE evolution, and tidal evolution. In contrast, the NS–LC population exhibits a wide range of eccentricities. For short-period ($\porb/\yr\le10$) detached \nslc\  binaries (which are particularly relevant to \gaia, given the observation timeline), we find that $\sim 27\%$ of \nslc\  binaries have highly eccentric orbits ($Ecc\ge0.5$). Between $57$–$59\%$ of these highly eccentric \nslc\ binaries contain NSs formed via CCSNe, the rest come from ECSNe or AIC. NSs formed via ECSNe dominate the overall \nslc\ binary population and typically contribute to low- to moderate-eccentricity orbits because of the adopted lower natal kicks \citep[\autoref{sec:SN-phy}][]{Podsiadlowski_2004,Ivanova_2018}. For example, $\approx82-83\%$ of \nslc\ binaries with nearly circular orbits ($\ecc\le0.1$) contain a NS that formed via ECSNe or AIC. For BHs, we adopt a fallback modulated kick distribution \citep[e.g.,][]{Belczynski_2012}. In both the \rapid\ and \delayed\ populations we find a wide range in eccentricities for the detached \bhlc\ binaries with $\porb/\yr\le10$. Although a significant fraction of the \bhlc\ binaries exhibit high eccentricities immediately post BH formation, binary interactions including tides reduce these eccentricities over time to create a significant low-eccentricity population at present day. In general, the \delayed\ population exhibits a higher fraction of \bhlc\ binaries with significant eccentricities due to typically stronger natal kicks compared to the rapid prescription. For example, $\sim30\%$ ($60\%$) of the detached \bhlc\ binaries with $\porb/\yr\le10$ are found in orbits with $\ecc\ge0.1$ in the \rapid\ (\delayed) population.  

The $\mlc$ distribution spans a wide range, extending up to $80\,\msun$ (\autoref{fig:intrinsic-mlc}). However, the maximum companion mass depends on the type of CO. For WD–LC binaries, $\mlc$ typically does not exceed $15\,\msun$, but can extend to much higher values. The actual $\mlc$ upper limit is significantly influenced by the number and nature of MT episodes during the course of their evolution. When MT is stable and occurs early, the LC can gain a significant amount of mass ($1$–$6\,\msun$) from the WD progenitor leading to rejuvenation and sometimes mass ratio reversal. For \nslc\ binaries, the upper limit extends to $\mlc/\msun\sim45$–-$50$, depending on the SN model, whereas, for \bhlc\ binaries, $\mlc$ can reach up to $80\,\msun$. The NS/\bhlc\ binaries with massive companions, $\mlc/\msun\ge40$, are of particular interest. They typically originate from systems where the NS/BH progenitor underwent a MT episode via RLOF leaving behind a high-mass, rejuvenated LC. Detection of such systems in the \gaia\ CO–LC population would be extremely valuable for probing the effects of mass accretion on CO companions \citep[][]{Neo_1977, Hellings_1983, Dray_2007, Renzo_2023}. Furthermore, a fraction of these binaries may be progenitors of future CO-CO binaries interesting for gravitational wave observations. 

The \colc\ population also shows a wide spread in the metallicity distribution $10^{-4}\le\metal\le0.03$ \footnote{The apparent pile-up at $\metal\approx0.03$ is not real. BSE's prescriptions are limited to $\metal\le0.03$. \cosmic\ assigns $\metal=0.03$ to any binaries with higher $\metal$ (see \autoref{sec:numerical setup}). } (\autoref{fig:intrinsic-met}). The estimated age, metallicity, and $\mlc$ of each observed \colc\ binary should help constrain the age and metallicity of its progenitor. Combined with mass measurements via astrometry or radial velocity followup, such constraints spanning a large range in metallicities may become very useful to constrain metallicity-dependent stellar evolution models \citep[][]{Klencki_2020}. Interestingly, our models predict that the relative abundance of different types of \colc\ binaries varies with metallicity. In particular, the relative fraction of \bhlc\ binaries declines sharply at higher metallicities while the fractions of \nslc\ and \wdlc\ binaries do not exhibit significant change \citep[][]{Mapelli_2013, Kinugawa2018, Iorio_2024, Nagarajan_2025b}. This trend can be understood in light of two physical processes. Mass loss from stellar winds, which inhibits core growth and results in a larger fraction of WDs and NSs instead of BHs, increases with metallicity \citep[][]{Belczynski_2010b, Yusof_2013}. On the other hand, lower metallicities lead to lower expansion of stellar radii, which in turn lead to suppression of binary interactions such as MT and CE \citep[e.g.,][]{Klencki_2020}. The short-period ($\porb/\yr\le10$) detached \bhlc\ binaries predominantly ($\ge90\%$) originate from CE evolution \citep[][]{Chawla2022}. Thus, at lower metallicities, suppression of CE evolution leads to the formation of a higher fraction of \bhlc\ binaries with wider orbits. We indeed find that the relative abundances of short-period ($\porb/\yr\le10$) and wider \colc\ binaries are significantly different for \bhlc\ binaries compared to the other types (\autoref{fig:intrinsic-mco}). Of course, these are competing effects. Thus, future detections of \colc\ binaries, in particular, the relative abundances of different types of COs as a function of metallicity will will be very interesting to investigate. 

\subsection{Gaia-resolvable population of CO--LC binaries}
\label{sec:Gaia-obs-pop}
\begin{figure*}[hbt!]
    \plotone{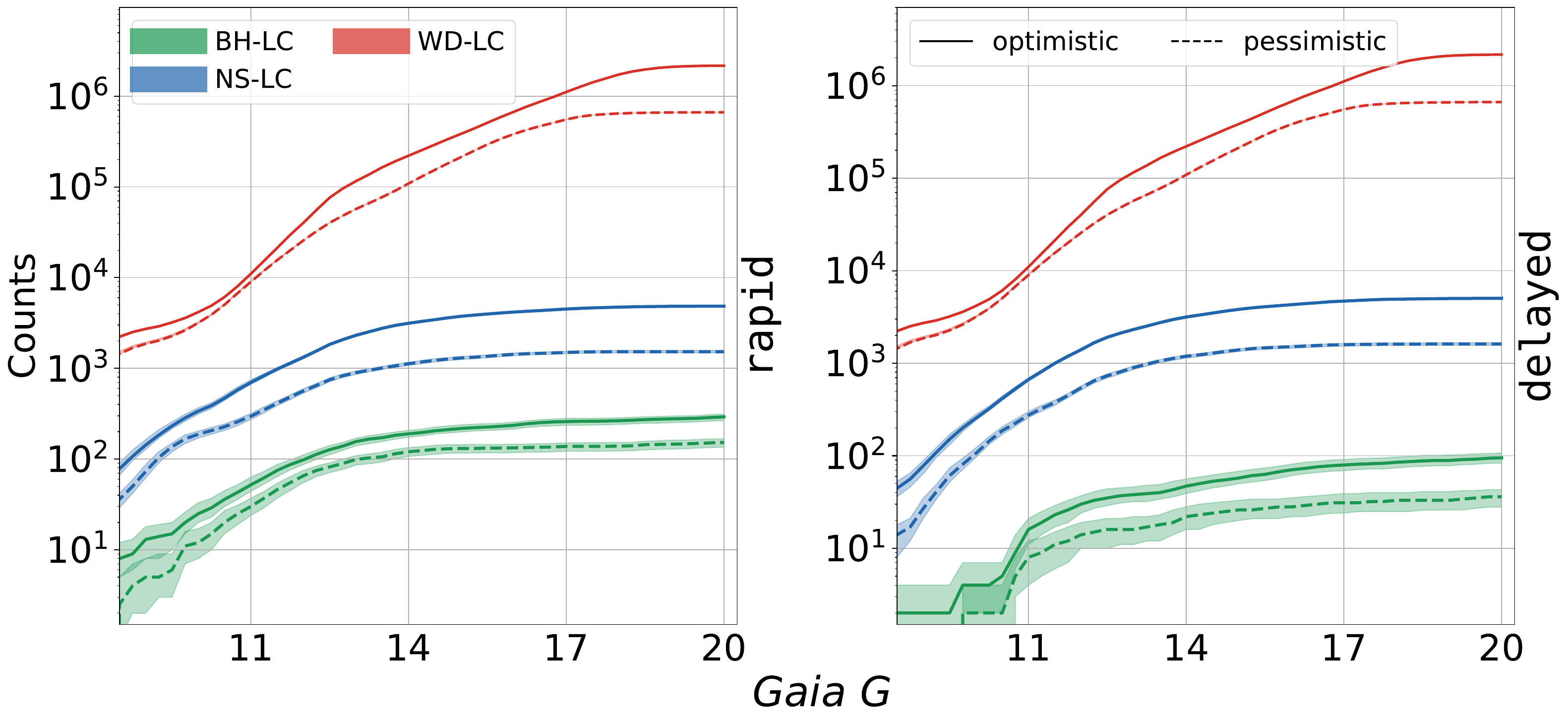}
    \caption{Cumulative distribution of \gaia\ resolved \colc\ population as a function of \gaia's G magnitude. Green, blue, and red color represent \bhlc, \nslc, and \wdlc\ population respectively. Solid and dashed curves represent the median of the number of detection for optimistic and pessimistic astrometric cuts, respectively. Shaded region represents the 10th and 90th percentile of the number of detection arising from the multiple galactic realisations (see \autoref{sec:Synthetic MW}).}
    \label{fig:detection_vs_gaiag}
\end{figure*}

\begin{figure*}
    \plotone{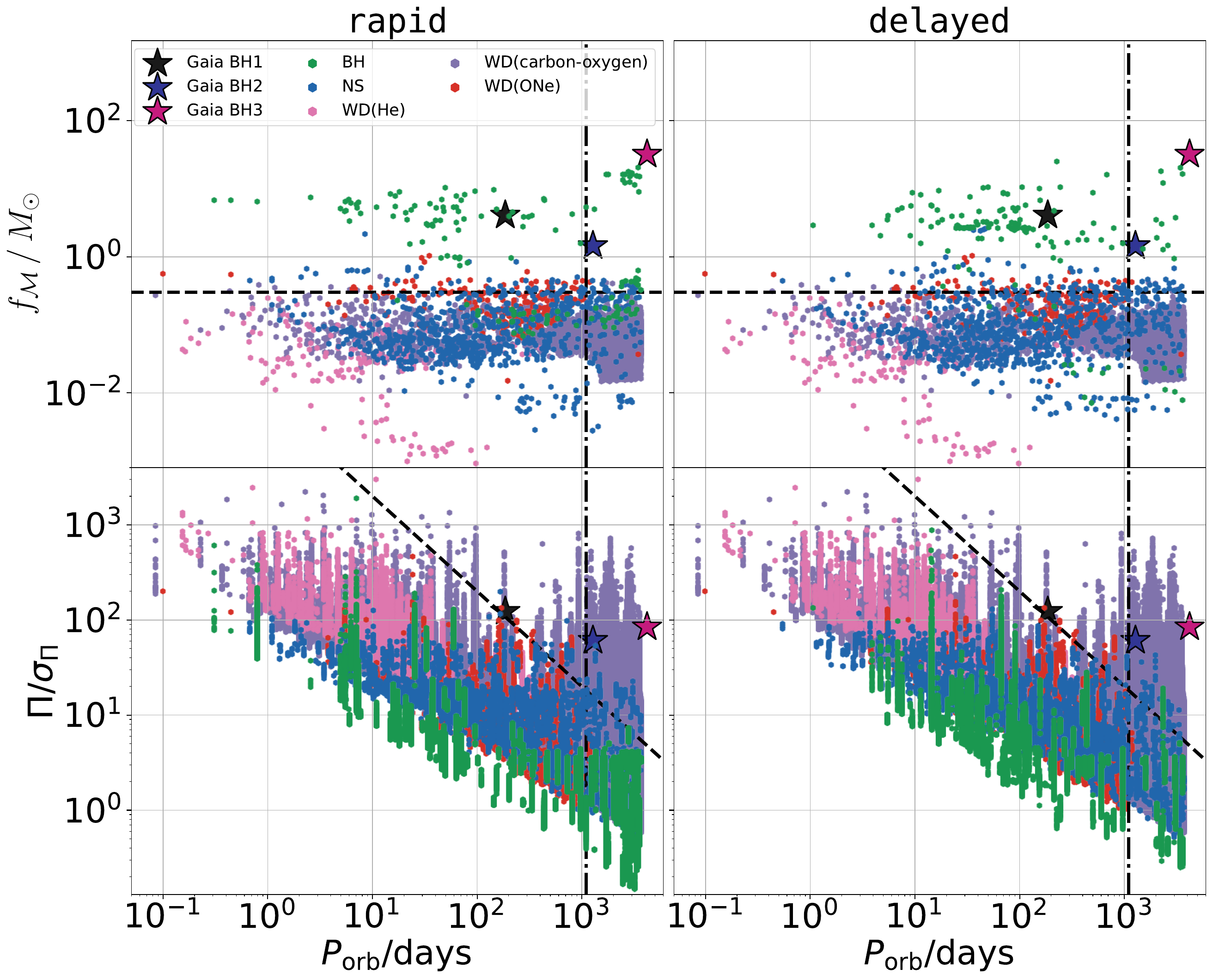}
    \caption{Distribution of parallax significance (bottom) and astrometric mass function ($f_{\mathcal{M}}$, top) as a function of $\porb$ for \gaia-resolvable \colc\ binaries in the \rapid\ and \delayed\ model. The vertical dashed line indicates the selection cut $\porb/\yr \le 3$, based on the DR3 observation duration. The bottom panel also includes the additional selection cut $\frac{\Pi}{\sigma_{\Pi}} \ge \frac{20000}{\porb}$ (dashed line), introduced in the DR3 pipeline to remove spurious binary star candidates with high $f_{\mathcal{M}} > 0.3\,\msun$ from the \nss\ catalog. This selection cut eliminates {\em all} potentially detectable \bhlc\ binaries in our models, retaining only NS and WD binaries. In the top panel, the horizontal dashed line marks $f_{\mathcal{M}} = 0.3\,\msun$, showing that a major fraction of \bhlc\ binaries lie above it.}
    \label{fig:porb_vs_snr_para}
\end{figure*}

\autoref{fig:detection_vs_gaiag} shows the cumulative distribution of the \gaia\ resolved \colc\ binaries as a function of \gaia's G magnitude. Our modeled populations show that over its 10-year observation period, \gaia\ would be able to resolve  $292^{+21}_{-26}\ (95^{+12}_{-12})$ \bhlc, $4843^{+71}_{-87}\ (5021^{+89}_{-74})$ \nslc\ and $\sim2\times10^{6}$ \wdlc\ binaries for the optimistic selection cuts for the \rapid\ (\delayed) model. Similarly, the numbers of detection with the more stringent pessimistic cuts are $152^{+14}_{-17}\ (36^{+7}_{-8})$, $1525^{+49}_{-47}\ (1612^{+52}_{-48})$, and $\sim6\times10^{5}$ for \bhlc, \nslc, and \wdlc\ binaries, respectively in our \rapid\ (\delayed) model.

These significant detection rates clearly indicate \gaia's potential to identify and characterize the \colc\ binaries in the MW. Binaries detected through astrometric measurements with \gaia\ will be complementary to those detected via X-ray \citep[][]{Corral-Santana2016} and radio observations for the BH, NS, and WD binaries. Such comprehensive detection strategies will not only increase the overall number of known systems but will also enable more detailed population studies, helping to constrain uncertainties in binary interaction physics and SN physics involved in the formation and evolution of \colc\ binaries \citep[][]{Langer_2020, Wang_2024}. 

The additional cut in DR3 based on the parallax significance, $\frac{\Pi}{\sigma_{\Pi}}$, adopted to filter out potentially spurious orbital solutions arising from the \gaia\ scanning law (see \autoref{sec:detection-criteria-dr3}), has a very strong effect on what is detectable in DR3. While the condition $\frac{\Pi}{\sigma_{\Pi}}\ge \frac{20000}{\porb}$ does not directly imply a threshold on $f_{\mathcal{M}}$, effectively, $f_{\mathcal{M}} > 0.3\,\msun$ solutions preferentially populate the region of spurious solutions. In \autoref{fig:porb_vs_snr_para} we show $\frac{\Pi}{\sigma_{\Pi}}$ and the astrometric mass function, $f_{\mathcal{M}}\equiv\frac{\mco^3}{(\mco+\mlc)^2}$, as a function of $\porb$ for all \colc\ binaries resolvable by \gaia\ over a 10-year observation period. The condition $\frac{\Pi}{\sigma_{\Pi}}\ge \frac{20000}{\porb}$ removes {\em all} \bhlc\ binaries and the majority of \nslc\ binaries in our model, although they are expected to be resolvable using the EOM cuts.

\begin{figure*}
    \plotone{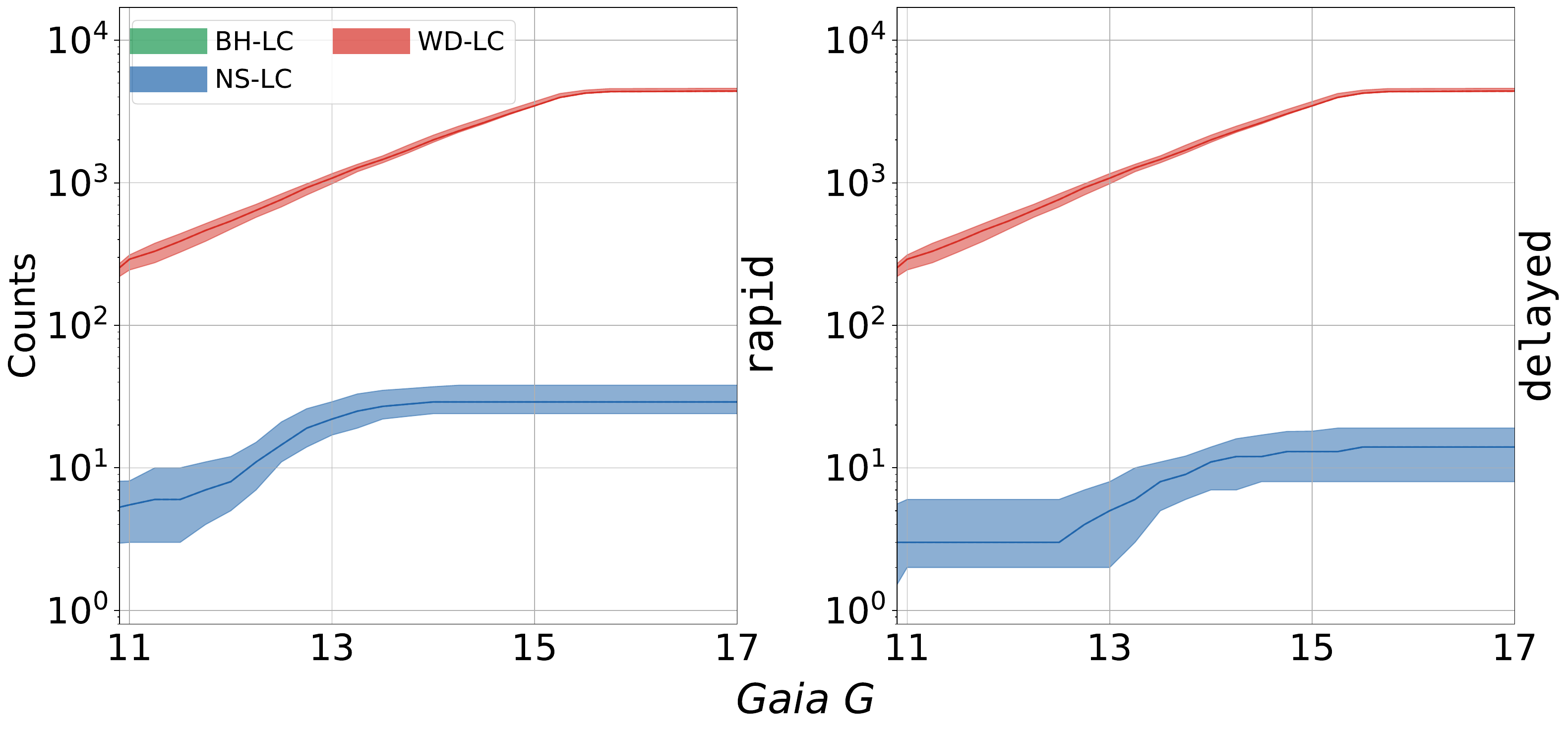}
    \caption{Same as \autoref{fig:detection_vs_gaiag} but for the \colc\ population after including the DR3 selection cuts (see \autoref{sec:detection-criteria-dr3}). The optimistic and pessimistic datasets merge in case of DR3-resolvable population because the DR3-resolvable binaries are wide enough that the entire optimistic dataset also satisfies $\alpha\ge3\sigma_{G}$.}
    \label{fig:detection_vs_gaiag_dr3}
\end{figure*}

\autoref{fig:detection_vs_gaiag_dr3} shows the cumulative distributions of detectable \colc\ binaries after incorporating DR3's additional cuts. In DR3, the expected number of \gaia\ resolvable \bhlc\ binaries reduces to zero, the number of \nslc\ binaries reduces to $20$--$40$, and the number of resolvable \wdlc\ binaries reduces to $\sim 4100-4300$. This is in reasonable agreement with the observed detached \colc\ candidate binaries identified using DR3, including three BHs \citep[][]{Andrews_2022a, El-badry2022e,El-badry_2023,Chakrabarti_2023,Tanikawa_2023,Panuzzo_2024}, approximately 21 NSs \citep[][]{El-badry_2024a,El-badry_2024b}, and around $\sim3200$ WDs \citep[][]{Shahaf_2022, Shahaf_2024}. Hence, based on our models, we expect only \wdlc\ and \nslc\ binaries to be present in the DR3 catalog. Notably, this result also highlights why modeling Gaia BH1- and Gaia BH2-like binaries through isolated binary evolution (IBE) remains challenging \citep[][]{El-badry2022e, El-badry_2023}.

The removal of all \bhlc\ binaries by the DR3 cuts reflects two compounding physical effects rather than any artifact of our population synthesis models. First, while the \bhlc\ population generated using \cosmic\ spans a continuous $f_{\mathcal{M}}$ distribution (\autoref{fig:f_m_vs_porb}), the subset surviving the \gaia\ EOM selection cuts is predominantly confined to $f_{\mathcal{M}} \gtrsim 0.3\,\msun$ (\autoref{fig:porb_vs_snr_para}). This concentration is driven by \gaia\ astrometry rather than by \cosmic: a \gaia-detectable binary preferentially has a wider separation, which increases the photocentric semi-major axis and hence the detection probability \citep[][]{Chawla_2023}. In the limit of negligible BH optical flux ($\mathcal{F}_{\rm BH} \ll \mathcal{F}_{\rm LC}$), the astrometric mass function reduces to $f_{\mathcal{M}} \sim M_{\rm BH}$, since the BH dominates the total mass of the binary \citep{Halbwachs_2023}. For example, a $3\,\msun$ BH orbiting a $1\,\msun$ LC yields $f_{\mathcal{M}} \approx 1.7\,\msun$, increasing to $\approx 3.5\,\msun$ and $\approx 8.3\,\msun$ for BH masses of $5\,\msun$ and $10\,\msun$, respectively. Astrometric mass functions of the order of a few solar-mass are thus the natural expectation for \bhlc\ binaries with typical companion masses, placing them inevitably in the $f_{\mathcal{M}} \gtrsim 0.3\,\msun$ regime. Second, the DR3 pipeline imposes a parallax significance cut to suppress spurious astrometric solutions arising from scan-angle-dependent signals in the \gaia\ data (\autoref{sec:detection-criteria-dr3}). These spurious solutions preferentially populate the $f_{\mathcal{M}} \gtrsim 0.3\,\msun$ regime --- precisely the region occupied by genuine \bhlc\ binaries --- because the scan-angle bias produces semi-major axes decoupled from Kepler's third law, yielding unphysically large $f_{\mathcal{M}}$ values. The DR3 reliability cuts therefore eliminate the spurious solutions and the \gaia\ EOM-detectable \bhlc\ population simultaneously. The three known \gaia\ BH systems (BH1, BH2, BH3; \citealt{El-badry2022e, El-badry_2023, Panuzzo_2024}), also shown in \autoref{fig:porb_vs_snr_para}, all fall outside these cuts --- consistent with their identification through targeted searches that bypassed the standard \nss\ pipeline rather than the blind DR3 selection; we discuss their properties and the implications for our models in \autoref{sec:detected BHs}.

\section{Comparison with detected \colc\ candidates}
\label{sec:comparison with CO--LC candidates}

\gaia's \nss\ catalog has already enabled the discoveries of several \colc\ candidates 
using astrometry, spectroscopy, and photometry. In the following sections, we provide an 
overview of the detected \colc\ candidates and compare the properties of the detected 
\colc\ candidates with our modeled populations where applicable.

\subsection{Detached \bhlc\ candidates discovered in DR3}
\label{sec:detected BHs}

Three detached \bhlc\ binaries have been identified using DR3 astrometry and radial velocity follow-up, referred to as \gaia\ BH1, BH2, and BH3 \citep[][]{El-badry2022e, El-badry_2023, Panuzzo_2024}. \gaia\ BH1 contains a $9.62\,\msun$ BH with a bright solar-type MS companion with $\mlc/\msun=0.93$ and $[{\rm Fe/H}]=-0.2$, in an orbit with $\porb/\days=185.6$ and $\ecc\approx0.45$. Theoretical modeling of the galactic orbit indicates a thin disk origin \citep[][]{El-badry2022e}. \gaia\ BH2 consists of an $8.9\,\msun$ BH orbiting a red giant with $\mlc/\msun\approx1$ in a fairly wide ($\porb/\days=1276$), eccentric ($\ecc\approx0.52$) orbit. \gaia\ BH3 is a $32.7\,\msun$ BH with a metal-poor ($[{\rm Fe/H}]=-2.56$) companion of mass $0.76\,\msun$ in a wide ($\porb/\days\sim4253$) eccentric ($\ecc\approx0.7$) orbit. Interestingly, only \gaia\ BH1 would marginally survive the DR3 cuts (\autoref{sec:detection-criteria-dr3}). None of the \bhlc\ binaries produced in our models survive the DR3 cuts.

\begin{figure}
    \plotone{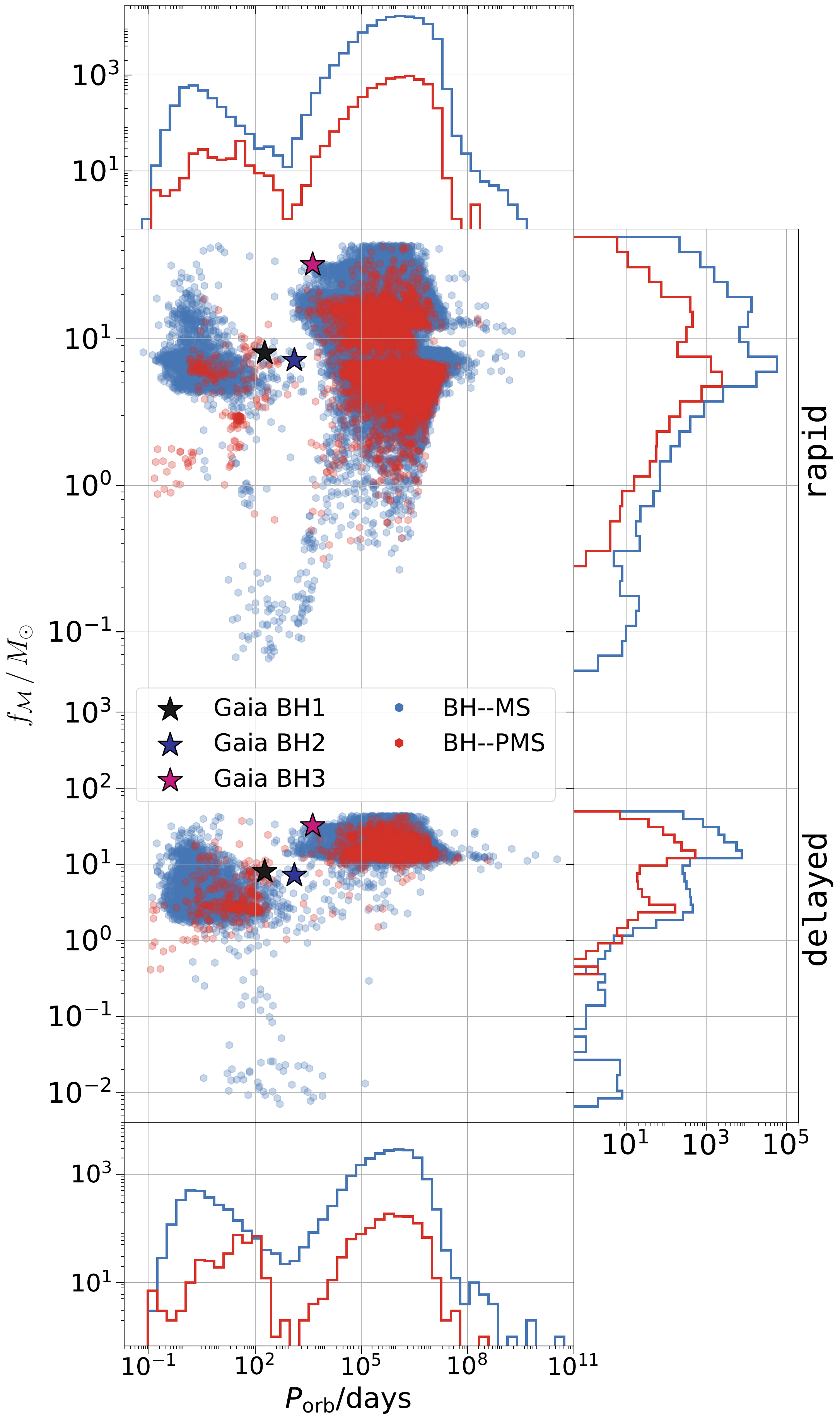}
    \caption{Distribution of astrometric mass function ($f_{M}$) as a function of $\porb$ 
    for \bhlc\ binaries in the \rapid\ and \delayed\ models.}
    \label{fig:f_m_vs_porb}
\end{figure}

As we dig deeper into the comparison using all \gaia-resolvable \colc\ binaries, we find some differences between the properties of \gaia\ BH1, BH2, and BH3 and the population of \bhlc\ binaries produced in our models. While the $f_\mathcal{M}$ in our models are in rough agreement with the observed systems, the model \bhlc\ binaries populate a very different region in the $\porb$ vs $\frac{\Pi}{\sigma_{\Pi}}$ plane (\autoref{fig:porb_vs_snr_para}), primarily because the parallax significance does not match. In \autoref{fig:f_m_vs_porb}, we compare the astrometric mass function ($f_\mathcal{M}$) vs $\porb$ of the simulated \bhlc\ population with the \gaia\ BHs. All three observed dormant BHs lie in the dip in the $\porb$ distribution of the dormant \bhlc\ binary population, caused by the combined effect of the CE phase and SNe kicks \citep[][]{Chawla2022}. This tension between the observed systems and our predictions suggests that the formation of \gaia\ BH1, BH2, and BH3 may not be fully explained by standard IBE.

In our simulated populations of intrinsic \bhlc\ binaries, we find that only $\sim0.04$--$0.3\%$ have orbital periods in the range $100$--$300$ days. Of these, about $4$--$5\%$ host BHs with $8\le\mbh/\msun\le10$ and $\mlc/\msun\sim1$, which we refer to as \gaia\ BH1-like candidates. In the \rapid\ model, none of these candidates have a solar-type LC; instead, they have sub-solar metallicities ($[{\rm Fe/H}]<-0.5$) and, unlike \gaia\ BH1, highly eccentric ($\ecc\sim0.9$) orbits. Their typical evolutionary pathway involves a massive primary ($20 \le \mprim/\msun \le 35$) and a low-mass ($< 1\,\msun$) secondary in an initially wide orbit with $\porb \sim 10$--$20$ yr. These systems undergo a single CE phase, which shrinks the orbit to $\sim 1$ day; following the SN of the primary, the orbit expands to $\sim 200-240$ days. Consequently, none of these candidates reproduce the observed stellar and orbital properties of \gaia\ BH1. In the \delayed\ model, however, we do find a \gaia\ BH1-like candidate with a solar-mass LC ($\mlc/\msun\sim0.8$, $[{\rm Fe/H}]\sim-0.15$) in an eccentric orbit ($\ecc\sim0.9$). This system starts with a massive primary ($M_\mathrm{pri}\sim71\,\msun$) and a low-mass companion in a $20\,\yr$ wide orbit; it undergoes a CE episode which shrinks its $\porb$ from $\sim3000$ days to $\sim1$ day, before the SN expands it to a present-day orbit of $106$ days with a kick of $\sim90\,\kms$. The kick is high enough to widen the orbit but not sufficient to reach the observed period of \gaia\ BH1.

\citet[][]{El-badry2022e} showed that \gaia\ BH1 cannot be produced through the IBE channel under standard CE assumptions; reproducing its orbital properties requires artificially enforcing an extreme CE efficiency of $\alpha_{\rm CE} \approx 14$. However, such a high value of $\alpha_{\rm CE}$ implies the presence of dominant additional heating mechanisms to eject the envelope during the CE phase and is therefore not expected \citep[e.g.,][]{Ivanova_2018}. In \citet[][]{Chawla2022}, we found that the CE channel is a major contributor to \gaia-resolvable \bhlc\ binaries, with $\sim 50\%$ of all resolvable \bhlc\ binaries forming through CE evolution. Notably, the contribution from CE evolution in the intrinsic BH--LC population with $\porb/\yr \le 10$ is even higher ($\sim 96\%$). Considering the resolvable BH--MS and BH--PMS populations separately, approximately $40\%$ ($95\%$) of resolvable BH--MS (BH--PMS) binaries are expected to have undergone at least one CE phase. Nevertheless, while CE evolution is common among dormant \bhlc\ binaries, our models show that it does not provide a viable formation pathway for \gaia\ BH1 under standard assumptions (\autoref{sec:numerical setup}).

Our simulated population also does not quite match all the properties of the \gaia\ BH2 system. In the BH--PMS population, only about $0.02\%$ of binaries have orbital periods between $1000$ and $1500$ days and these systems host \bhlc\ binaries with $\mbh\sim6\,\msun$ orbiting solar-mass companions. Thus, the BHs in our simulated population is not as massive as \gaia\ BH2. On the other hand, if we apply selection cuts requiring $8\le\mbh/\msun\le10$, PMS companion, and $0.8\leq\mlc/\msun\leq1.5$, the surviving systems fall into two distinct categories. One group occupies tight ($\porb\le100$ days) circular orbits, typically as a result of post-SN MT episodes. The other group resides in very wide orbits with periods longer than $10^{5}$ days and moderate eccentricities ($\ecc\approx0.5$--$0.7$), having experienced no MT episodes. This bimodality in the present-day orbital-period distribution reflects two distinct formation channels: one involving multiple stable and unstable mass-transfer episodes and another that avoids mass transfer entirely, but none producing a \gaia\ BH2-like binary quite accurately for both the \rapid\ and \delayed\ SN models. This is in agreement with \citet[][]{El-badry_2023} who find that the formation of \gaia\ BH2 through a common-envelope channel is challenging.

Several theoretical studies have attempted to explain the formation of \gaia\ BH1 and BH2 through modifications to the standard assumptions of the IBE channel, including alternate CE prescriptions \citep[][]{El-badry2022e, El-badry_2023, Olejak_2025, Mapelli_2026}, triple system evolution \citep[][]{Hayashi_2023, Generozov_2024, Li_2024, Li_2026, Naoz_2025, Tanikawa_2025}, modifications to the maximum radius reached by the BH progenitor during the giant phase \citep[][]{Gilkis_2024}, fine-tuning of the kick magnitude and orientation \citep[][]{Kotko_2024}, and modified stellar wind prescriptions that avoid mass transfer or CE evolution entirely \citep[][]{Kruckow_2024}. While most of these studies find it challenging to reproduce the observed properties of \gaia\ BH1 and BH2 under standard assumptions of IBE, there is no consensus on the required formation channel.

Alternatively, such \bhlc\ binaries could be produced inside young star clusters \citep[YSCs, e.g.,][]{Portegies_Zwart_2010, Shikauchi2020, Banerjee_2026}, dense star clusters \citep[][]{Chatterjee2017, Fantoccoli_2025} and disrupted dense star clusters \citep[][]{Schiebelbein_2026} via dynamical processes. Recent numerical studies have shown the potential of YSCs to form \gaia\ BH1 and BH2-like binaries \citep{Rastello_2023, Tanikawa_2023a, Di_Carlo_2024}, indicating that YSCs with initial mass $\sim3\times10^2$--$3\times10^4\,\msun$ are likely most efficient in their production.

Our simulated BH--LC population produces only $\sim0.07$--$0.6\%$ \gaia\ BH3-like binaries. These systems host BHs with $\mbh\sim31\,\msun$ orbiting sub-solar mass companions ($\mlc\sim0.7$--$0.9\,\msun$) in eccentric ($0.2\le\ecc\le0.9$) but extremely wide orbits ($\porb\sim300$--$7000\,\yr$), unlike the observed \gaia\ BH3 system. The evolution of progenitors of \gaia\ BH3-like binaries in our models start with a massive primary ($M_\mathrm{pri}\ge35\,\msun$) with a sub-solar mass companion ($\mlc\sim0.7$--$0.9\,\msun$) in an eccentric orbit ($0.2\le\ecc\le0.9$) with initial orbital periods ($\ge300\,\yr$), which do not experience any MT episode during their evolution. However, if we relax the the $\mbh$ range to be between $21$--$30\,\msun$, we obtain about $\sim0.04$--$0.1\%$ \bhlc\ binaries with $\mlc/\msun \sim 0.8$, in an orbit with $5500\le\porb/{\rm days}\le6500$ and moderate eccentricity ($0.1$--$0.4$). None of these binaries experience an MT episode, like the ones with more massive BHs ($\mbh/\msun\sim31$) in wider orbits ($\porb/\yr\sim300-7000$). Our models thus produce BH3-like binaries in terms of mass, $\porb$, and eccentricity, consistent with the findings of \citet{Iorio_2024}, who showed that \gaia\ BH3-like \bhlc\ binaries formed via IBE also never undergo any mass-transfer episodes, with the observed properties primarily set by the ZAMS properties and the details of SN physics. While our models do produce \gaia\ BH3-like binaries, albeit with slightly less massive BH companions ($\mbh/\msun\le30$), this demonstrates that such systems can in principle form through IBE. The fact that we do not recover a closer analog to \gaia\ BH3 is likely a consequence of the limited resolution of our simulations; the higher-resolution setup of \citet[][]{Iorio_2024} was able to reproduce a system more closely resembling \gaia\ BH3 in its stellar and orbital architecture \citep[see also][]{El-badry_2024c}.

Beyond IBE, \gaia\ BH3 may also have originated through dynamical processes. The Galactic orbit of \gaia\ BH3 and the chemical composition of its LC establish its association with the ED-2 stream, believed to be created by a disrupted globular cluster of mass between $2\times10^3$--$4\times10^4\,\msun$ \citep[][]{Balbinot_2023, Balbinot_2024, Dodd_2023}, suggesting a dynamical formation channel \citep[e.g.,][]{Marin_2024, Marin_2026}. However, a detailed chemical analysis by \citet[][]{Hackshaw_2025} found no chemical peculiarities in the red giant companion of \gaia\ BH3, with its chemical normalcy consistent with both dynamical capture and IBE, leaving the formation pathway of this system still to be determined.

Overall, \gaia's astrometrically detected \bhlc\ binary population remains an evolving story. The relatively low number of \bhlc\ binaries identified in DR3 is almost certainly due to DR3-specific selection biases, with a strong expectation of significant detections in future data releases. Most recently, \citet{Nagarajan_2025a} showed that while the number of observed \bhlc\ binaries in DR3 is somewhat larger compared to what is expected from IBE, it is significantly lower compared to the predictions of the dynamical channels. We expect that future data releases would be very interesting to resolve these issues.

\subsection{Detached \nslc\  candidates discovered in DR3}

\begin{figure}
    \plotone{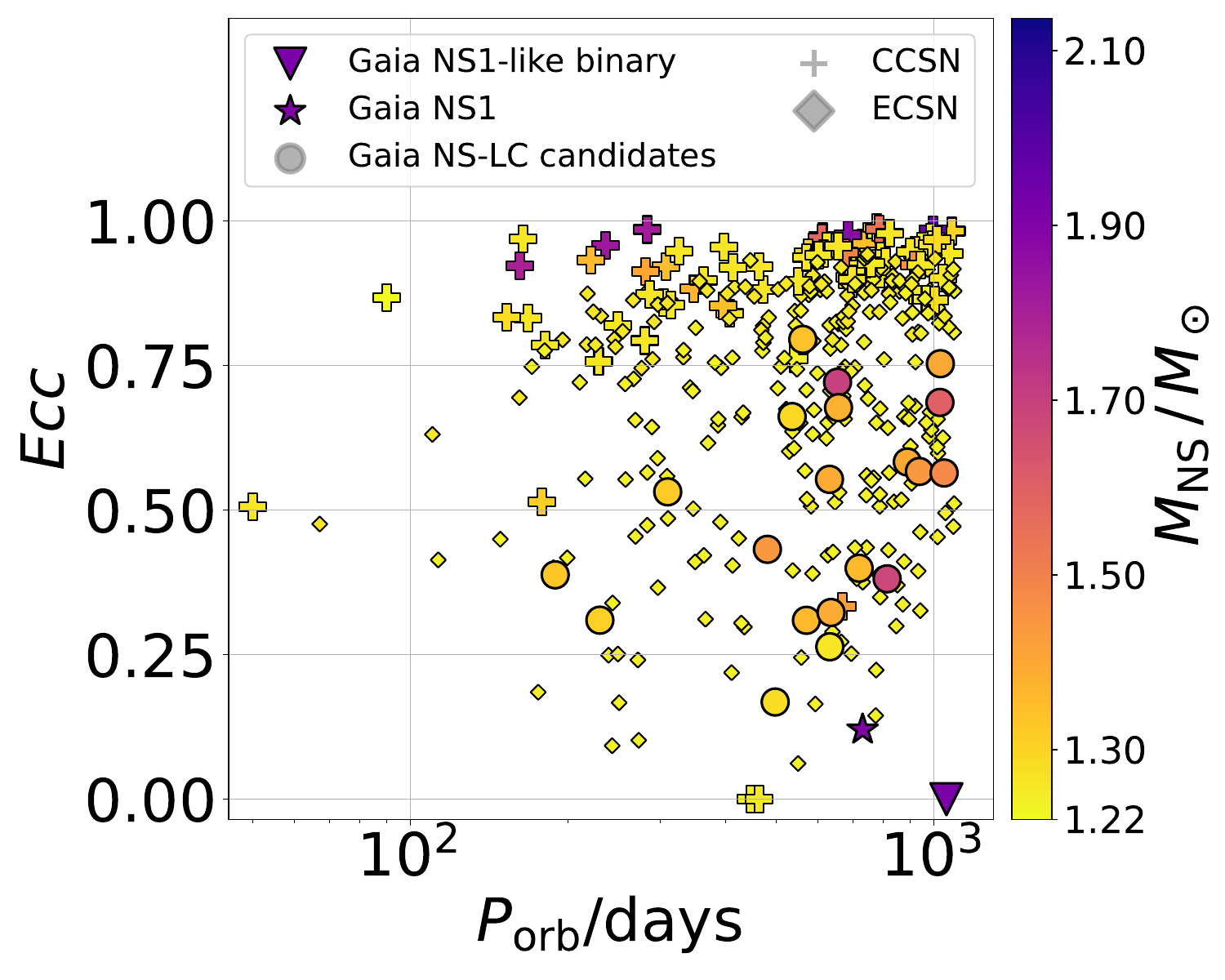}
    \caption{Distribution of the $\porb$ vs Ecc for the DR3 resolvable \nslc\ binaries in our simulated population, compared with the \gaia-detected NS–LC candidates. Plus and cross markers represent modeled NS–LC binaries where the NS is formed via CCSN and ECSN, respectively, while dots denote the \gaia\ observed NS–LC candidates from \citet[][]{El-badry_2024b}. The triangle and star represent the \gaia\,NS1-like candidate from our simulated data and the actual \gaia\,NS1 from \citet[][]{El-badry_2024a}, respectively.}
    \label{fig:dr3_theory_vs_obs_NS}
\end{figure}

\citet[][]{El-badry_2024b} have reported 21 \nslc\ candidates in the DR3 catalog. For these observed \nslc\ candidates, the $\mns$ distribution ranges between $1.25-1.9\,\msun$, with MS companions with $\mlc$ ranging between $0.7-1.3\,\msun$. They have near-solar metallicities ($-0.5\le[{\rm Fe/H}]\le0.5$) and are within $1.5\,\rm{kpc}$. Most of the \nslc\ binaries have $\porb$ between $10^{2}-10^{3}$ days and inferred eccentricity higher than $0.4$. 

Our models predict roughly $20$--$40$ resolvable \nslc\ binaries in DR3. These have $\mns/\msun\sim1.25$--$1.9$. The $\porb$ is between $10^{2}$--$10^{3}$ days with a wide eccentricity spread $0$--$0.98$ (\autoref{fig:dr3_theory_vs_obs_NS}). About $\sim90\%$ of the \nslc\ binaries with NSs formed via CCSN are highly eccentric ($\ecc\ge0.8$); in contrast, those produced via ECSN exhibit a more uniform spread in eccentricity. Clearly, our modeled \nslc\ properties and those of the observed candidates show reasonably good agreement; however, the $\mns$ for the modeled population are slightly lower than \gaia's \nslc\ candidates. Since a major fraction of the parameter space occupied by \gaia\ \nslc\ candidates overlaps with NSs modeled via ECSN, the discrepancy in $\mns$ is almost certainly a consequence of our assumed lower limit on the He-core masses that collapse to NSs via ECSN (see \autoref{sec:SN-phy}). This is supported by previous studies \citep[e.g.,][]{Linden_2009, Andrews_2015}, which find that a higher He-core mass range ($2$--$2.5\,\msun$) for ECSN better reproduces the Galactic double neutron star population and the high-mass X-ray binary population in the Small Magellanic Cloud.

Within our modeled DR3 population, we identify one \nslc\ binary whose properties closely match those of \gaia\ NS1 \citep[][]{El-badry_2024a}. This system hosts a $1.9\,\msun$ NS with a $0.89\,\msun$ companion in a circular orbit with $\porb=1056$ days (pink star in \autoref{fig:dr3_theory_vs_obs_NS}). Its formation proceeds through several distinct phases. The NS progenitor expands during the giant branch and initiates a CE phase, which shrinks the orbit from $674$ days to $\sim0.5$ days and fully circularizes it. At NS formation, the natal kick widens the orbit to $\sim2.4$ days and re-introduces eccentricity. Subsequently, the companion fills its Roche lobe and transfers mass to the NS, expanding the orbit to its present-day separation before the binary detaches; this final mass-transfer episode also re-circularizes the orbit. The resulting system has a present-day age of $\approx9\,\gyr$, comparable to the inferred age of \gaia\ NS1. This demonstrates that, although high-eccentricity \nslc\ binaries dominate the population, DR3-detectable \nslc\ binaries with circular or near-circular orbits can also form through a post-supernova mass-transfer episode.

\subsection{Detached \wdlc\ candidates discovered in DR3}
\label{sec:WD_DR3}
\begin{figure*}
    \plotone{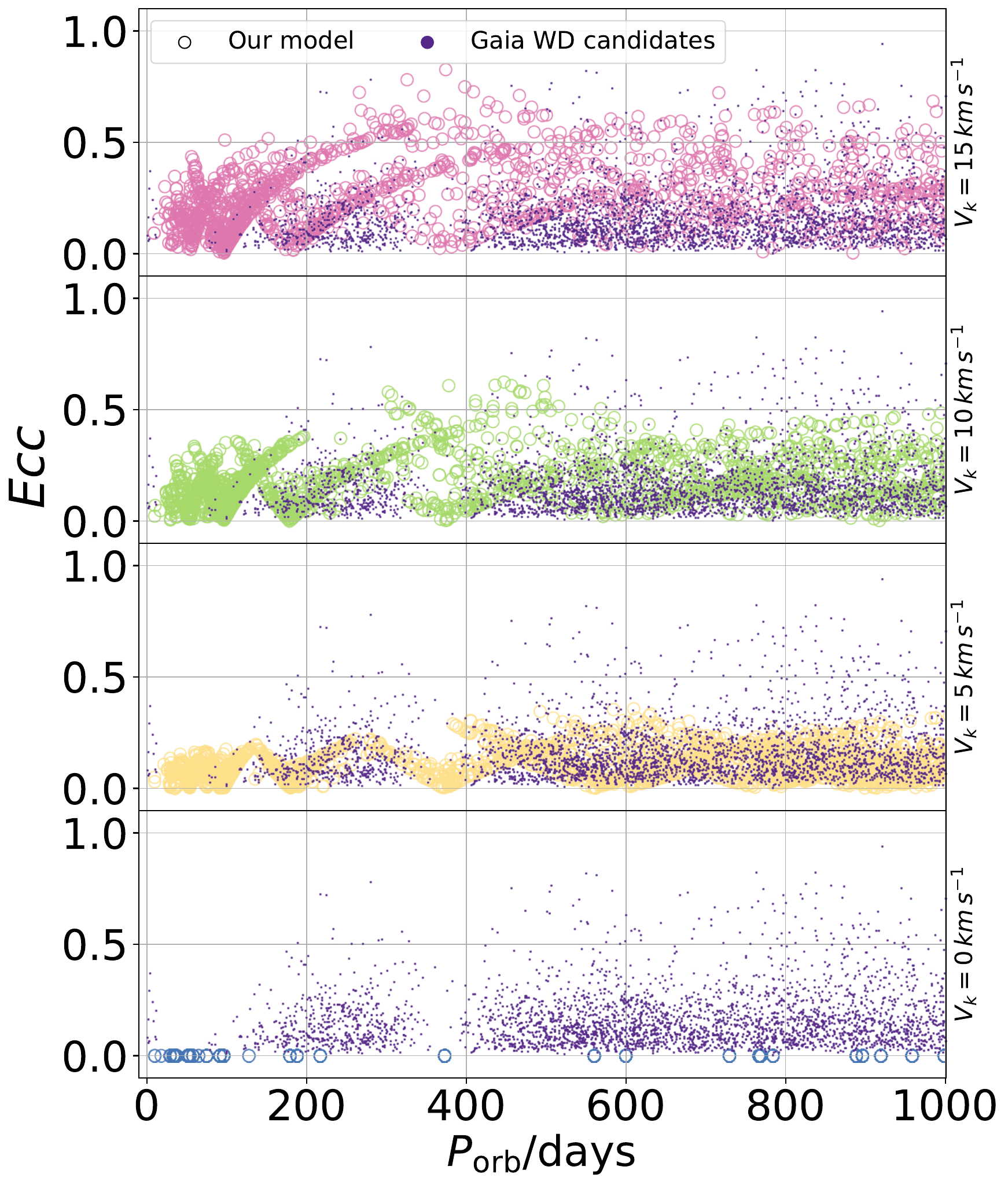}
    \caption{Distribution of the $\porb$ vs $\ecc$ for the DR3 resolvable \wdlc\ binaries in our simulated populations compared with the \gaia\ \wdlc\ candidates \citep[][]{Shahaf_2022, Shahaf_2024, Ganguly_2023} candidates.}
    \label{fig:WD_ecc_vs_porb}
\end{figure*}
\citet[][]{Shahaf_2022, Shahaf_2024} identified $\sim3200$ \wdlc\ candidates using DR3 astrometry and photometry. \citet{Ganguly_2023} further analyzed the \colc\ candidates identified by \citet{Andrews_2022a,Shahaf_2022} using spectral energy densities to properly classify the nature of the COs and constrain the stellar properties of the COs and their LCs. The categorization between NSs and WDs by \citet{Shahaf_2022} was done based on a simplistic clustering algorithm in the $\mco$--$\ecc$ plane. Followup study by \citet{Ganguly_2023} found that several COs categorized as NSs by \citet{Shahaf_2022} show significant UV excess and can best be explained if these sources contain WDs and not NSs. Similar misclassifications were also found by \citet{El-badry_2024b} using radial-velocity followup. A large number of \wdlc\ binaries have also been identified by \citet[e.g.,][]{Nayak2024} using spectral energy density analysis of multi-wavelength observed fluxes from a wide array of surveys including \gaia\ and GALEX, although, the orbital properties are unknown for these sources. Taking into account the findings of all of these studies we proceed to compare the properties of \wdlc\ binaries identified using DR3 astrometry to those in our models. 

The sources identified by \citet{Shahaf_2022, Shahaf_2024}, corrected by followup studies by \citet{Ganguly_2023} have all properties of interest for this study. These observed \wdlc\ binaries exhibit $0.3\le\mwd/\msun\le0.8$, $\porb/\days\le10^3$, and $0<\ecc<0.9$. 

Our model predicts $\sim4300$ \gaia-detectable \wdlc\ binaries in the DR3 catalog, comparable to the $\sim3200$ reported in the analyses of \citet[][]{Shahaf_2022,Shahaf_2024}. Notably, whereas the \wdlc\ candidates in \gaia's \nss\ catalog exhibit a range of orbital eccentricities, our modeled \wdlc\ population is dominated by circular orbits. This discrepancy arises because we impose an upper $\porb$ cutoff of 3 years for DR3 observations. Consequently, most of these binaries are expected to have undergone at least one mass-transfer episode, which circularizes their orbits.

To investigate this further, we perform a simple experiment by assuming fixed natal-kick magnitudes ($V_k$) of $5$, $10$, and $15\,\kms$ during WD formation, and examine their effects on the $\ecc$ of DR3-resolvable \wdlc\ binaries. For each fixed magnitude, we apply isotropically oriented kicks, compute the resulting eccentricities and $\porb$, and then apply the \gaia\ selection biases (see \autoref{sec:detection-criteria}). We find that $V_{k}$ values in the range $5$–$15\,\kms$ increase the eccentricities of the \gaia-detectable \wdlc\ population. At the same time, these moderate kicks disrupt a significant fraction ($10$–$40\%$) of binaries, depending on the magnitude and orientation of the kick. \autoref{fig:WD_ecc_vs_porb} compares our modeled population with the \wdlc\ candidates \citep[][]{Shahaf_2022, Shahaf_2024, Ganguly_2023} from \gaia's \nss\ catalog \citep[][]{Arenou_2023}, demonstrating that the assumed natal kicks produce a population that aligns well with \gaia\ observations in the $Ecc-\porb$
parameter space. 

Hydrodynamic simulations indicate that asymmetric mass ejection during CE evolution can impart kicks of order $3$--$4\ \kms$ at WD birth \citep[][]{Sandquist_1998}. Such natal kicks have been invoked across a wide range of contexts: the deficit of WDs in nearby open clusters \citep[][]{Kalirai_2001, Fellhauer_2003}; the spatial distribution and velocity dispersion of young WDs in globular clusters \citep[][]{Davis_2006, Davis_2008, Heyl_2007a, Heyl_2007b, Heyl_2008a, Heyl_2008b, Calamida_2008, Grondin_2024}; and the delayed core collapse of globular clusters\citep[][]{Fregeau_2009, Heyl_2009, Richer_2009}. They have likewise been proposed to account for the rotation rates of single WDs \citep[][]{Spruit_1998}, the orbital properties of wide WD--LC binaries \citep[][]{Izzard_2010, El-badry_2018, Connor_2025, Hwang_2025}, and the evolution of WD triples \citep[][]{Hamers_2019, Shariat_2023, Shariat_2025}. More speculatively, natal kicks may influence planetary systems---for instance, by contributing to the formation of hot Jupiters in WD binaries \citep[][]{Stephan_2024}, free-floating planets \citep[][]{Stehan_2026} and to the tidal disruption of planetesimals that drives the metal pollution of WD atmospheres \citep[][]{Stone_2015, Connor_2023, Akiba_2024}.

The WD kick indeed provides a plausible way to reproduce the observed WD--LC parameter space; however, other mechanisms such as introducing additional recombination energy during the CE phase or CE evolution proceeding during the thermally pulsating AGB phase of the WD progenitor also offer reasonable explanations for the observed wide eccentric \wdlc\ binaries in the \nss\ catalog \citep[e.g.][]{Yamaguchi_2024b, Yamaguchi_2025}. While our results are robust within the assumed kick velocities, future work will explore their sensitivity to a broader range of kick scenarios.

\section{Conclusions} \label{sec:conclusion}

In this work, we present a detailed analysis of the \gaia-detectable detached \colc\ populations. We discuss the key present-day observable properties of the intrinsic \colc\ binary populations in the Milky Way and investigate the same for the subsets that are resolvable by \gaia's astrometry with DR3 and EOM selection cuts.

\begin{itemize}
    \item The model intrinsic present-day Milky Way population of \colc\ binaries exhibits a wide spread in the $\mco$ distribution, spanning from $0.1\,\msun$ to $45\,\msun$. WDs, NSs, and BHs exhibit mass ranges $0.1\le\mwd/\msun\le1.4$, $1.2\le\mns/\msun\le3$, and $3\le\mbh/\msun\le45$ (\autoref{fig:intrinsic-mco}).
    \item The $\mco$ distribution (\autoref{fig:intrinsic-mco}) clearly distinguishes the populations generated by the two adopted SN models. The \rapid\ SN prescription yields a pronounced lower mass-gap ($3-5\,\msun$) in the $\mco$ distribution, whereas the \delayed\ prescription does not exhibit such a gap. Notably, a small fraction ($\sim 1\%$) of BHs located within the mass gap in the \rapid\ model form via AIC of NSs.
    \item The intrinsic \nslc\ binaries with $\porb/\yr\le10$ in our models do not contain NSs with $\mns/\msun\ge2$. We find that all of these high-mass NSs with short-period orbits are accreting via RLOF. This indicates that such high mass NSs are unlikely to be found in detached systems by \gaia. Indeed, no such candidate \nslc\ binaries have been identified using DR3 yet. 
    \item We find a significant difference in the $\ecc$ distribution for \colc\ binary orbits depending on the adopted SN prescription (\autoref{fig:intrinsic-ecc}). The \delayed\ model exhibits higher $\ecc$ compared to the \rapid\ model; $\sim70\%$ ($30\%$) of \bhlc\ binaries with $\porb/\yr\le10$ exhibit $\ecc\le0.1$ in the \rapid\ (\delayed) model. 
    \item All model \colc\ binaries detectable by \gaia's DR3 or EOM go through MT, CE, and tidal evolution, all of which contribute to erasing the initial orbital $\ecc$. The final $\ecc$ is set by natal kicks during CO formation. As a result, devoid of natal kicks for WDs, the detectable \wdlc\ binaries produced in our fiducial model exhibit only circular orbits. In contrast, with large natal kicks, $57\%$ ($18\%$) of the \nslc\ binaries with NSs formed via CCSN have $\ecc>0.5$ ($\ecc<0.1$). Due to the adopted lower natal kicks for ECSN, $\sim42\%$ ($40\%$) of the \nslc\ binaries with NSs formed via ECSN have $\ecc>0.5$ ($\ecc<0.1$) (see \autoref{sec:intrinsic-pop}).       
    \item We find that the relative production of \colc\ binaries is metallicity dependent: systems with a BH as the compact object decline in frequency at higher metallicities due to enhanced stellar-wind mass loss, whereas the formation rate of WD/NS--LC binaries remains largely unaffected (see \autoref{fig:intrinsic-met}).
    \item Our models predict that by EOM, \gaia\ could identify $10^2$--$10^3$ \bhlc, $10^3$--$10^4$ \nslc, and $\sim10^6$ \wdlc\ binaries using astrometry alone. The respective expected numbers in DR3 from our model are zero \bhlc, $20$--$40$ \nslc, and $\sim4\times10^3$ \wdlc\ binaries (see \autoref{sec:Gaia-obs-pop}).   
    \item The predicted number as well as the orbital properties of DR3-detectable \nslc\ binaries in our models agree well with the observed \nslc\ binaries (see \autoref{fig:dr3_theory_vs_obs_NS}). 
    \item The orbital properties, in particular the observed eccentricities for the \wdlc\ binaries require modest $5$--$15\,\kmps$ natal kicks during WD formation (see \autoref{fig:WD_ecc_vs_porb}). 
\end{itemize}

In \citet[][]{Chawla2022,Chawla_2023}, we demonstrated the potential of the \gaia\ sky survey to detect the \bhlc\ binary population using astrometric, photometric, and spectroscopic observations. This work extends our previous efforts and complements our earlier findings, further establishing our prediction that \gaia\ is expected revolutionize the study of \colc\ binaries through its discoveries \citep[][]{Andrews_2022a, El-badry2022e, El-badry_2023, El-badry_2024a, El-badry_2024b, Shahaf_2022, Shahaf_2024, Chakrabarti_2023, Tanikawa_2023, Panuzzo_2024}. It is heartening that our earlier predictions are roughly consistent with DR3 bolstering our belief that we are on the verge of detecting a large population of detached \colc\ binaries spanning wide ranges in age, metallicity, and $\mco$ \citep[][]{Lam_2026, Muller_2026, Simon_2026}. Future \gaia\ data releases, along with complementary surveys such as LSST \citep[][]{Ivezic_2019} and TESS \citep[][]{Ricker2014}, promise to refine our understanding of high-mass stellar evolution and CO formation even further \citep[][]{Korol_2017, Johnson_2018, Masuda_2019, Wiktorowicz2021, Wiktorowicz_2025, Breivik_2022, Sorabella_2022, Chawla_2023, Sajadian_2024, Yamaguchi_2024a, Green_2025, Nir_2025}. Furthermore, these observations are expected to enhance our insights into binary evolution \citep[][]{Langer_2020, Wang_2024, Schurmann_2025, Xu_2025, Parkosidis_2026, vanSon_2026}. A fraction of the detectable \colc\ binaries are expected to have properties commensurate to produce double-compact binaries in the future, thus provide input for potential progenitors of gravitational wave studies. 

\section*{Acknowledgement}

We thank the anonymous referee for their constructive comments, which helped improve the 
manuscript. CC acknowledges support from TIFR's graduate and IISER Rubin Observatory Fellowship. SC acknowledges support from the Department of Atomic Energy, Government of India, under project no. 12-R\&D-TFR-5.02-0200 and RTI 4002. All simulations were done using cloud computing on Azure. The Flatiron Institute is supported by the Simons Foundation.

\software{astropy \citep{astropy:2013, astropy:2018, Astropy_2022}
          \cosmic\ \citep{Breivik_2020, Breivik_2021}; \mwdust\ \citep{Bovy2016}; \texttt{isochrones}\ \citep{Isochrones_2015}; \texttt{matplotlib}\ \citep{matplotlib}; \texttt{numpy}\ \citep{numpy}; \texttt{scipy}\ \citep{SciPy-NMeth_2020}; \texttt{pandas}\ \citep[][]{Pandas_2010}}

\bibliography{references}{}
\bibliographystyle{aasjournal}

\end{document}